\documentclass[pra,aps]{revtex4}
\usepackage{graphicx}
\usepackage{colordvi}
\begin{document}
\title{Two-dimensional symbiotic solitons and quantum droplets in \\
a quasi-one-dimensional optical lattice}

\author{S. M. Al-Marzoug$^{1,3}$, B. B. Baizakov$^2$, and H. Bahlouli$^3$}
\affiliation{$^1$ Interdisciplinary Research Center for Intelligent
Secure Systems, King Fahd University of Petroleum and Minerals, Dhahran 31261, Saudi Arabia,\\
$^2$ Physical-Technical Institute, Uzbek Academy of Sciences,
100084, Tashkent, Uzbekistan,\\
$^3$ Physics Department, King Fahd University of Petroleum and
Minerals, Dhahran 31261, Saudi Arabia}
\date{\today}

\begin{abstract}
Symbiotic solitons (SS) and quantum droplets (QD) are self-trapped
localized modes emerging in binary Bose-gas mixtures with
intra-component repulsion and inter-component attraction. We have
shown that two-dimensional SS can be stabilized against collapse or
decay by means of a quasi-one-dimensional optical lattice (OL).
Mobility of SSs along the free direction of the potential allows us
to explore interactions and collisions of SSs moving in the same
channel and neighboring channels of the quasi-1D OL. For the case of
equal atom numbers in both components of the binary Bose-Einstein
condensate (BEC) we have developed a variational approach that
showed the stability of SS. For parameter settings when the SS stays
on the verge of collapse instability we take into account the
Lee-Huang-Yang quantum fluctuations (QF) term in the coupled
Gross-Pitaevskii equations. Repulsive QF prevents the mean-field
collapse and gives rise to formation of 2D QDs with peculiar
properties such as incompressibility and surface tension, which are
inherent to liquids. The proposed model of binary BEC loaded in a
quasi-1D OL allows us to demonstrate the manifestations of the
incompressibility and surface tension of 2D QDs. The formation of QD
in imbalanced binary BEC with different numbers of atoms of the
components in the presence of a quasi-1D OL has been investigated.
The possible application of the proposed model to studies of
anisotropic superfluidity is discussed.

\end{abstract}
\maketitle

\section{Introduction}

Interacting two-component Bose-gas mixtures can exhibit a variety of
interesting phenomena which are not observed in single-component
gases. Depending on the character (attractive or repulsive) and
strength of same-species and cross-species interactions such
phenomena as miscibility and phase separation, formation of
localized bound states, and other types of density-modulated
patterns, may show up in these systems. Theoretical prediction
\cite{petrov2015} and experimental observation
\cite{cabrera2018,semeghini2018} of a new state of matter emerging
in binary Bose-Einstein condensates (BEC), which became known as
{\it quantum droplet} initiated very active research aimed at
understanding its basic properties (see review articles
\cite{luo2021,guo2021} and book \cite{malomed-book-2022}).

Quantum droplets (QD) are extremely dilute (with atomic density
$\sim 10^{15}$ cm$^{-3}$) compared to ordinary liquids, such as
water (with molecular density $\sim 10^{22}$ cm$^{-3}$). Despite the
huge difference in atomic/molecular densities, QDs possess some
physical properties typical to ordinary liquids, such as
incompressibility and surface tension
\cite{petrov2018,alba-arroyo2022}. The self-binding property and
stability of multi-dimensional QDs in free space are the direct
manifestation of their surface tension. The liquid-like nature of
QDs was demonstrated in the experimental study of their binary
collisions \cite{ferioli2019}.

Another type of self-trapped localized state which can exist in
two-component Bose-gases with same-species repulsion and
cross-species attraction is so-called {\it symbiotic soliton}. These
localized modes are entirely stable in one-dimensional settings
\cite{adhikari2005,perezgarcia2005,adhikari2008}, while they are
prone to collapse instability and decay in multi-dimensional free
space~\cite{malomed-book-2022}. In order to stabilize 2D SS some
kind of external potential has to be employed. In particular,
stabilization of 2D symbiotic solitons by means of 2D optical
lattice (OL) was reported in \cite{ma2016}. Although the 2D OL
provides complete stability of 2D SSs, their mobility in these
conditions becomes restricted due to the Peierls-Nabarro type
barrier \cite{kivshar1993}. The possibility of solving the problem
of limited motion with the help of a quasi-1D OL was proposed for 2D
solitons in single component attractive BEC~\cite{baizakov2004}. The
proposed setting give the opportunity to explore the interactions
and collisions between 2D scalar solitons stabilized by a quasi-1D
OL. To the best of our knowledge, the issues related to stability
and dynamics of 2D symbiotic/vector solitons in quasi-1D OL has not
been addressed so far.

A common feature of SS and QD is that they both emerge in
two-component Bose-Einstein condensates with intra-component
repulsion and inter-component attraction, but are stabilized by
different physical factors. Namely, 2D and 3D QDs are stabilized by
repulsive quantum fluctuations and can exist in free space, while
SSs are stabilized by means of an external potential but cannot
exist in free space.

Multidimensional SS on the verge of collapse instability, when the
inter-component attraction slightly overcomes the intra-component
repulsion, represents a useful system to explore the collapse in
two-component media, as well as the effect of quantum fluctuations.
At this delicate balance between attractive and repulsive atomic
interactions, when the mean-field competing nonlinearities almost
cancel each other, the effect of relatively small quantum
fluctuations starts to play a decisive role. On the other hand, the
minimal strength of the external potential (e.g. quasi-1D OL)
capable of preventing the collapse instability in 2D SS can be
related to the value of the LHY term \cite{lee1957}, thus providing
indirectly a physical measure of quantum fluctuations.

Our objective in this paper consists in showing that 2D SS can be
stabilized by quasi-1D OL. The advantage of the present model over
previously reported ones is that, here 2D SS can move along the free
potential direction. This gives the opportunity to explore
interactions and collisions of SS and QDs moving in the same and
adjacent channels of the quasi-1D OL. We show that the dynamical
features of 2D SS in the present model significantly differ from
those of 2D scalar solitons, previously reported in
\cite{baizakov2004}. For parameter values, such that the mean field
focusing and de-focusing nonlinearities nearly balance each other,
we take into account the stabilizing effect of quantum fluctuations
in the governing equations. The main subject of interest in this
case will be the manifestation of incompressibility and surface
tension of QDs in presence of quasi-1D optical lattice.

The paper is organized as follows: in the next Sec. II the model
governing equations are presented and a variational approach for 2D
symbiotic solitons is developed. Sec. III is devoted to
manifestations of the surface tension of 2D quantum droplets in the
presence of a quasi-1D optical lattice. In Sec. IV we consider the
formation and dynamics of quantum droplets in imbalanced Bose-gas
mixtures with significantly different atom numbers of the
components. The main findings of this work are summarized in the
last Sec. V.

\section{Two-dimensional symbiotic solitons and variational approach}

Dynamics of a two-component BEC loaded in an external potential is
governed by the following dimensionless coupled GPE
\cite{stoof-book}
\begin{equation}\label{gpe}
i\partial_t\psi_i + \Delta\psi_i - V(x,y)\psi_i +
(g_{i}|\psi_i|^2+g_{ij}|\psi_j|^2)\psi_i =0, \quad i, j  = 1,2,
\quad i \not= j,
\end{equation}
where $\psi_i(x,y,t)$ are the mean field wave functions of the
components, $\Delta = \partial^2/\partial x^2 + \partial^2/\partial
y^2$, $V(x,y)$ is the external potential. Below we consider a
quasi-one-dimensional optical lattice potential $V(x,y) = V_0 \cos(k
y)$ of strength $V_0$ and wave vector $k$, and assume the
coefficients of intra-species and inter-species atomic interactions
to be repulsive ($g_1<0, g_2<0$) and attractive ($g_{12}=g_{21}
> 0$), respectively.

The stationary solutions of the above system are seeked in the
following forms $\psi_i(x,y,t) \rightarrow e^{-i\mu_i
t}\psi_i(x,y)$, with $\mu_i$ being the chemical potentials, obey the
following equations
\begin{equation}\label{us}
\Delta \psi_i - V(x,y)\psi_i - \psi_i^3 + g\,\psi_j^2\psi_i +
\mu_i\psi_i = 0, \quad i, j  = 1,2, \quad i \not= j,
\end{equation}
where we normalized our parameter settings such that
$g_{1}=g_{2}=-1$ and $g_{12}=g_{21}=g$. The Lagrangian that gives
rise to Eqs. (\ref{us}) is
\begin{equation}\label{lagr}
L=\frac{1}{2}\int_{-\infty}^{\infty}\left[(\nabla \psi_1)^2+(\nabla
\psi_2)^2+V(x,y) (\psi_1^2+\psi_2^2)+(\psi_1^4+\psi_2^4)/2 - g
\psi_1^2 \psi_2^2 - \mu_1 \psi_1^2 - \mu_2 \psi_2^2 \right] dx dy.
\end{equation}
The shape of a 2D localized mode stabilized by quasi-1D optical
lattice can be approximated using the following trial functions
\cite{baizakov2004}
\begin{equation}\label{ansatz}
\psi_i(x,y) = A_i \exp\left[-x^2/(2 a_i^2)-y^2/(2 b_i^2)\right],
\end{equation}
where $A_i,a_i,b_i$ denote the amplitude and widths of the component
$i=1,2$. The norm is given by $N_i=A_i^2 a_i b_i \pi$. Elongation in
the free direction $x$ of the quasi-1D OL corresponds to $a_i>b_i$.

Substitution of the ansatz (\ref{ansatz}) into Eq. (\ref{lagr}) and
spatial integration yields the effective Lagrangian
\begin{equation}\label{L}
2L = \sum_{i=1}^{2}\left[\frac{N_i}{2}\frac{a_i^2+b_i^2}{a_i^2
b_i^2} + N_i V_0\exp\left(-a_i^2 k^2/4\right) + \frac{N_i^2}{4\pi
a_i b_i}\right] - \frac{g N_1 N_2}{\pi \sqrt{(a_1^2 +
a_2^2)(b_1^2+b_2^2)}} - \mu_1 N_1 - \mu_2 N_2.
\end{equation}

The parameters of the localized mode can be found from the
variational equations $\partial L/\partial a_i=\partial L/\partial
b_i=0$, $i=1,2$.
\begin{eqnarray}
\frac{1}{a_1^3}&+&\frac{N_1}{4\pi a_1^2 b_1}-\frac{g N_2 a_1}{\pi
(a_1^2+a_2^2)^{3/2}(b_1^2+b_2^2)^{1/2}}+\frac{V_0 k^2}{2} a_1 \exp\left(-a_1^2 k^2/4\right)=0, \label{a1}\\
\frac{1}{b_1^3}&+&\frac{N_1}{4\pi a_1 b_1^2}-\frac{g N_2 b_1}{\pi
(a_1^2+a_2^2)^{1/2}(b_1^2+b_2^2)^{3/2}}=0, \label{b1}\\
\frac{1}{a_2^3}&+&\frac{N_2}{4\pi a_2^2 b_2}-\frac{g N_1 a_2}{\pi
(a_1^2+a_2^2)^{3/2}(b_1^2+b_2^2)^{1/2}}+\frac{V_0 k^2}{2} a_2 \exp\left(-a_2^2 k^2/4\right)=0, \label{a2}\\
\frac{1}{b_2^3}&+&\frac{N_2}{4\pi a_2 b_2^2}-\frac{g N_1 b_2}{\pi
(a_1^2+a_2^2)^{1/2}(b_1^2+b_2^2)^{3/2}}=0.\label{b2}
\end{eqnarray}
Solving numerically the above system of coupled algebraic equations
for given norms $N_i$, coefficient of inter-species attraction $g$
and parameters of the optical lattice $V_0, k$ one obtains the
widths of the localized mode ($a_i, b_i$). After defining the
amplitudes through the norms $A_i=\sqrt{N_i/(\pi a_i b_i)}$ it is
easy to construct the waveforms (\ref{ansatz}).

The chemical potentials $\mu_i$ are produced by the variational
equations $\partial L/\partial N_i=0$
\begin{equation}\label{mui}
\mu_i = \frac{a_i^2+b_i^2}{2 a_i^2 b_i^2}+V_0 \exp\left(-a_i^2
k^2/4\right)+\frac{N_i}{2\pi a_i b_i}-\frac{g
N_j}{\pi\sqrt{(a_i^2+a_j^2)(b_i^2+b_j^2)}}, \quad i,j = 1,2, \quad i
\not= j.
\end{equation}
For the symmetric case $N=N_1=N_2$, $a=a_1=a_2$, $b=b_1=b_2$
equations (\ref{a1})-(\ref{mui}) can be reduced to simpler form
\begin{eqnarray}
&&a^4\exp\left(-a^2 k^2/4\right)=\frac{2}{V_0 k^2}\left[\frac{(g-1)N a}{4\pi b}-1\right], \label{a} \\
&&b=\frac{4\pi a}{(g-1)N}, \label{b} \\
&&\mu=\frac{a^2+b^2}{2 a^2 b^2} + V_0 k^2 \exp\left[-a^2 k^2/4
\right]-\frac{(g-1)N}{2\pi a b}. \label{mu}
\end{eqnarray}
Since the left hand side of Eq. (\ref{a}) must be positive, from
Eqs. (\ref{a})-(\ref{b}) we get the condition for the existence of
2D symbiotic soliton in our model.
\begin{equation}
g > \frac{4\pi b}{N a}+1.
\end{equation}
This condition implies, that the strength of inter-component
attraction $g$ must be greater than the intra-component repulsion
(which has been set $g_{11}=g_{22}=1$) by additional quantity $4\pi
b/(N a)$.

\begin{figure}[htb]
\centerline{\centerline{ a) \hspace{4cm} b) \hspace{4cm} c)
\hspace{4cm} d)} \vspace{0.2cm}} \centerline{
\includegraphics[width=4cm,height=4cm,clip]{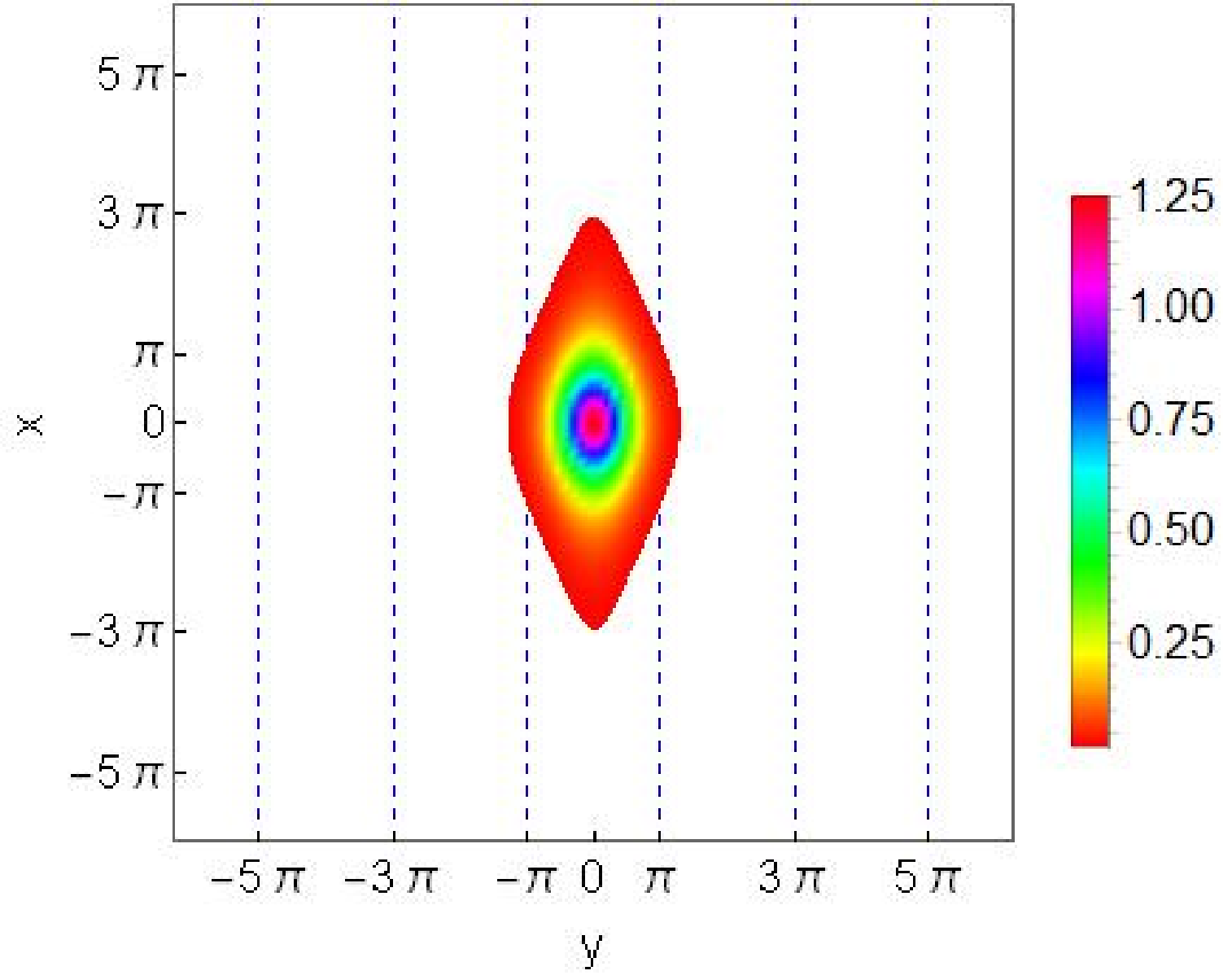}\quad
\includegraphics[width=4cm,height=4cm,clip]{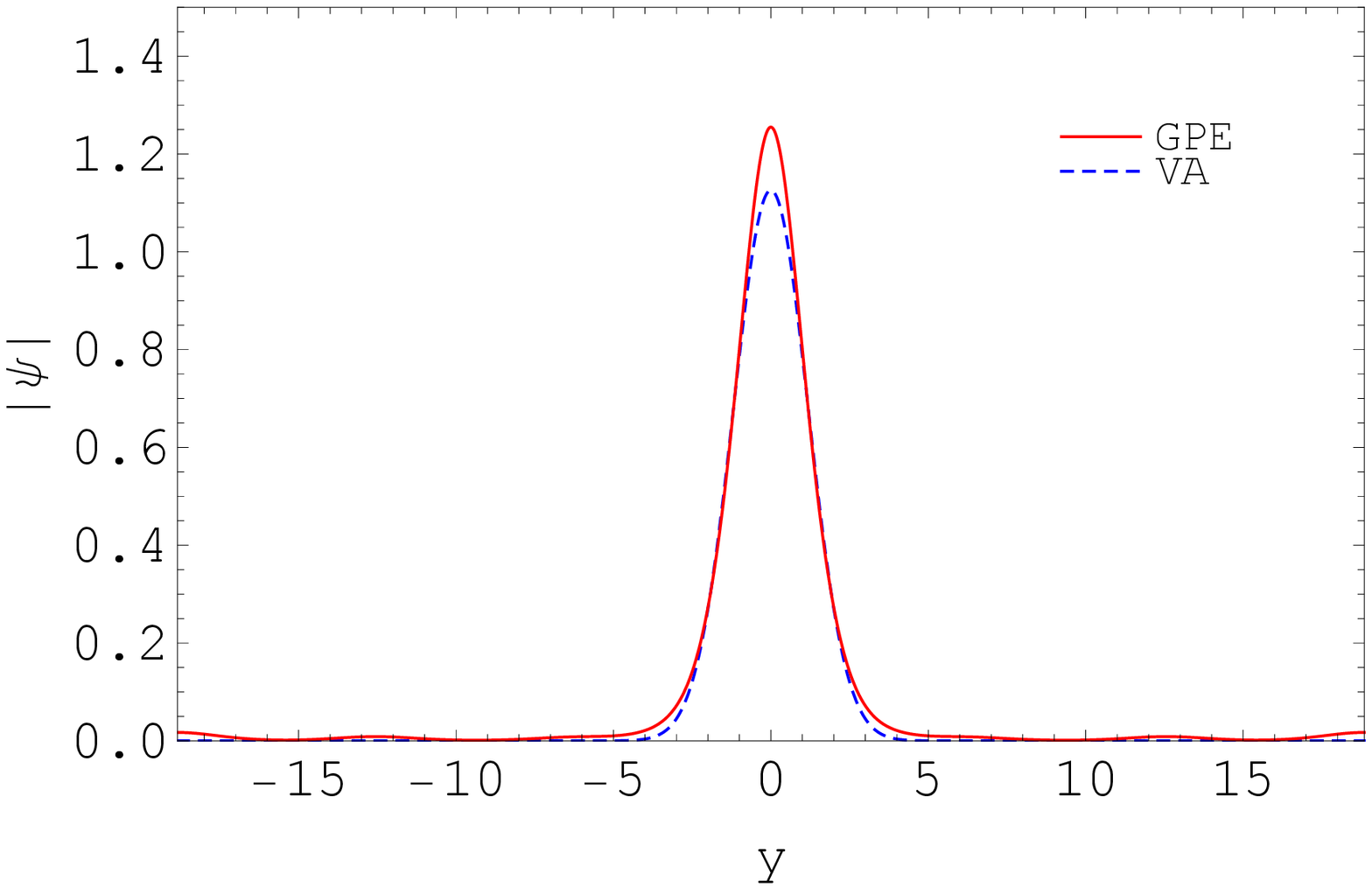}\quad
\includegraphics[width=4cm,height=4cm,clip]{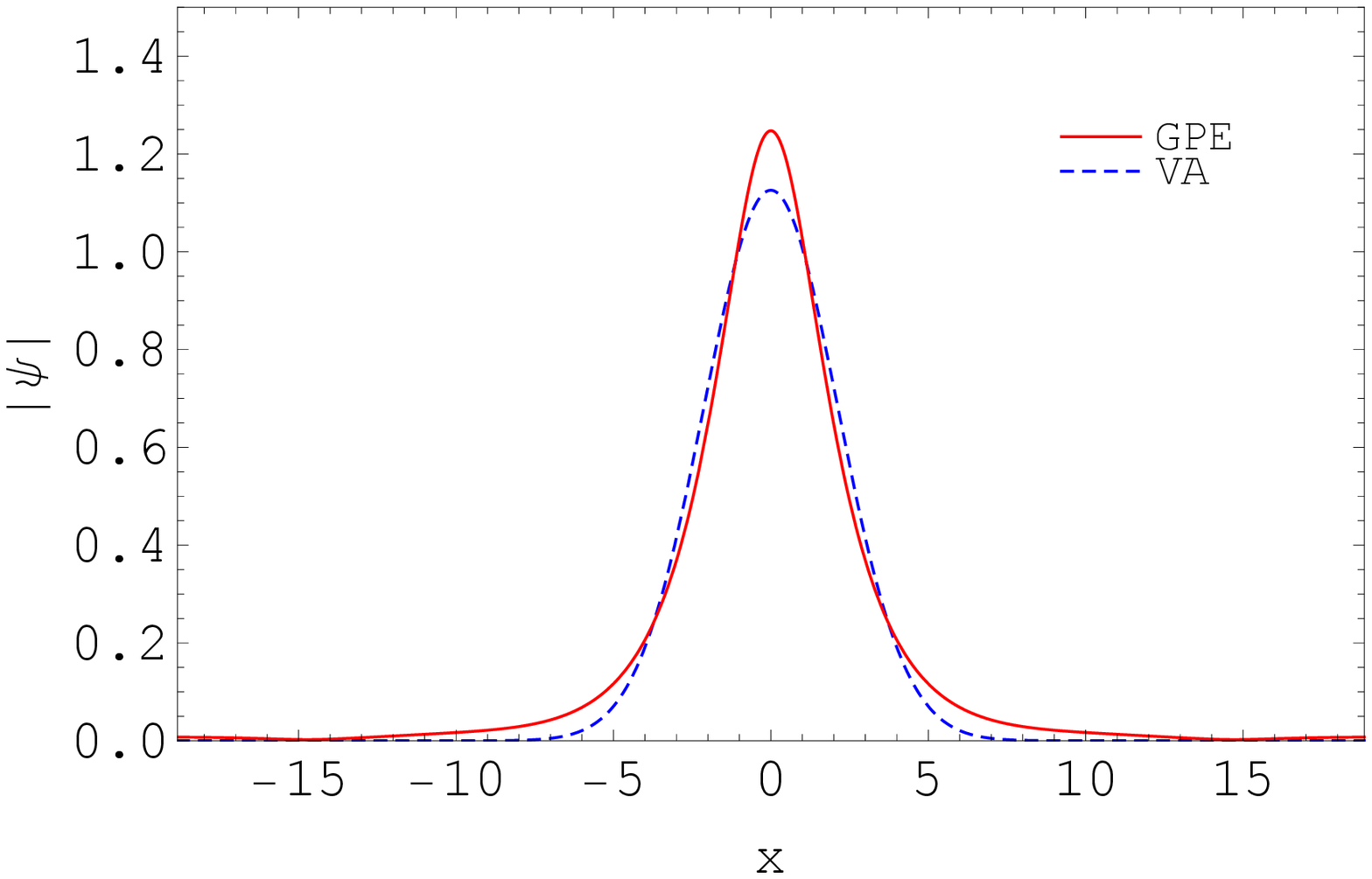}\quad
\includegraphics[width=4cm,height=4cm,clip]{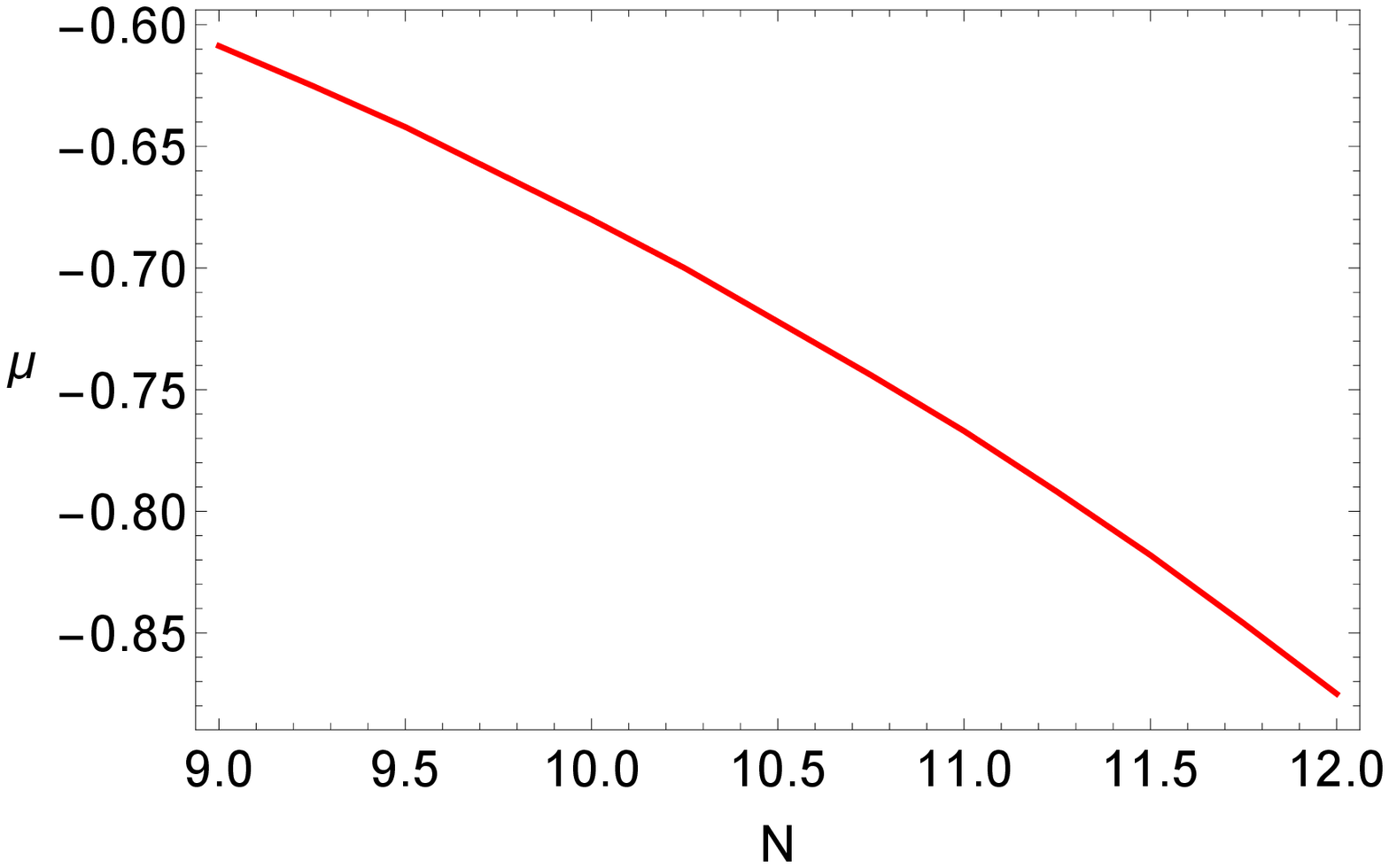}}
\caption{a) Density plot for the symmetric ($\psi = \psi_1 =
\psi_2$) 2D SS stabilized by a quasi-1D optical lattice $V(x,y)=V_0
\cos(k y)$. Vertical dashed lines indicate the maxima of the OL
potential. b) The wave profile $|\psi(0,y)|$ along the $y$-axis for
$x=0$, according to GPE (red line) and VA (blue dashed line). c) The
same for $|\psi(x,0)|$ along the $x$-axis for $y=0$. Strong
elongation of the wave profile in the free direction ($x$) is
evident. d) The chemical potential as a function of the norm showing
the negative slope $d \mu/d N < 0$. The prediction of VA Eqs.
(14)-(15): $A=1.13$, $a=1.18$, $b=2.12$. Parameter values: $V_0=-1$,
$k=1$, $N=10$, $g=1.7$. } \label{fig1}
\end{figure}

In Fig. \ref{fig1} we show the density plot and wave profiles of the
symmetric 2D SS soliton in a quasi-1D OL, whose parameters are found
from VA equations (\ref{a})-(\ref{b}), supplemented with the
expression $A=\sqrt{N/(\pi a b)}$. As can be seen from the middle
panels of this figure the VA well describes the peculiar shape of
the self-trapped mode in the form of a 2D Gaussian function,
elongated in the free direction of the quasi-1D OL.
The chemical potential as a function of the norm given by Eq.
(\ref{mu}) is shown in Fig. \ref{fig1}d, which gives supporting
evidence on the stability of the obtained solutions according to the
Vakhitov-Kolokolov criterion ($d \mu/dN < 0$) \cite{vakhitov1973}.
Numerical test of the stability was performed (but not shown) by
introducing weak random spatial modulation upon the wave profile of
the ground state solution and time evolution of the perturbed wave
function according to GPE(\ref{gpe}). We observed that SS sheds out
small amplitude linear waves and then restores its initial smooth
waveform, which is the indication of its stability.

The ground state wave profiles of the binary condensate in a
quasi-1D OL are obtained using the Pitaevskii phenomenological
damping procedure \cite{choi1998}, applied to coupled GPE
(\ref{gpe}). The numerical simulations are performed by the
split-step fast Fourier transform method using $256 \times 256$
Fourier modes \cite{agrawal-book,numrecipes} with a time step
$\Delta t=0.001$. The length of the integration domain was ${x,y}
\in [-8\pi, 8\pi]$ for producing the ground state wave profiles and
checking their stability, while ${x,y} \in [-12\pi, 12\pi]$ for
simulation of soliton collisions.

\subsection{Collisions of symbiotic solitons}

A quasi-1D OL not only stabilizes 2D SS against collapse or decay
but also gives the opportunity to explore their interactions and
collisions. Relevant numerical simulations have been performed in
two steps. At first, we produce two stable SS at a sufficiently
large distance from each other, either in the same channel, or
neighboring channels of the quasi-1D OL. Next, we apply a weak
harmonic potential ($\sim \alpha x^2$) in the free direction of the
OL, which induces the motion of solitons towards each other. When
the solitons are set in motion, the harmonic trap can be switched
off. The solitons moving in the same channel experience frontal
collisions, while those moving in the adjacent channels interact
through their overlapping tails.

\begin{figure}[htb]
\centerline{t = 8 \hspace{3cm} t = 11 \hspace{3cm} t = 14
\hspace{3cm} t = 17 \quad} \vspace{0.2cm} \centerline{
\includegraphics[width=4cm,height=4cm,clip]{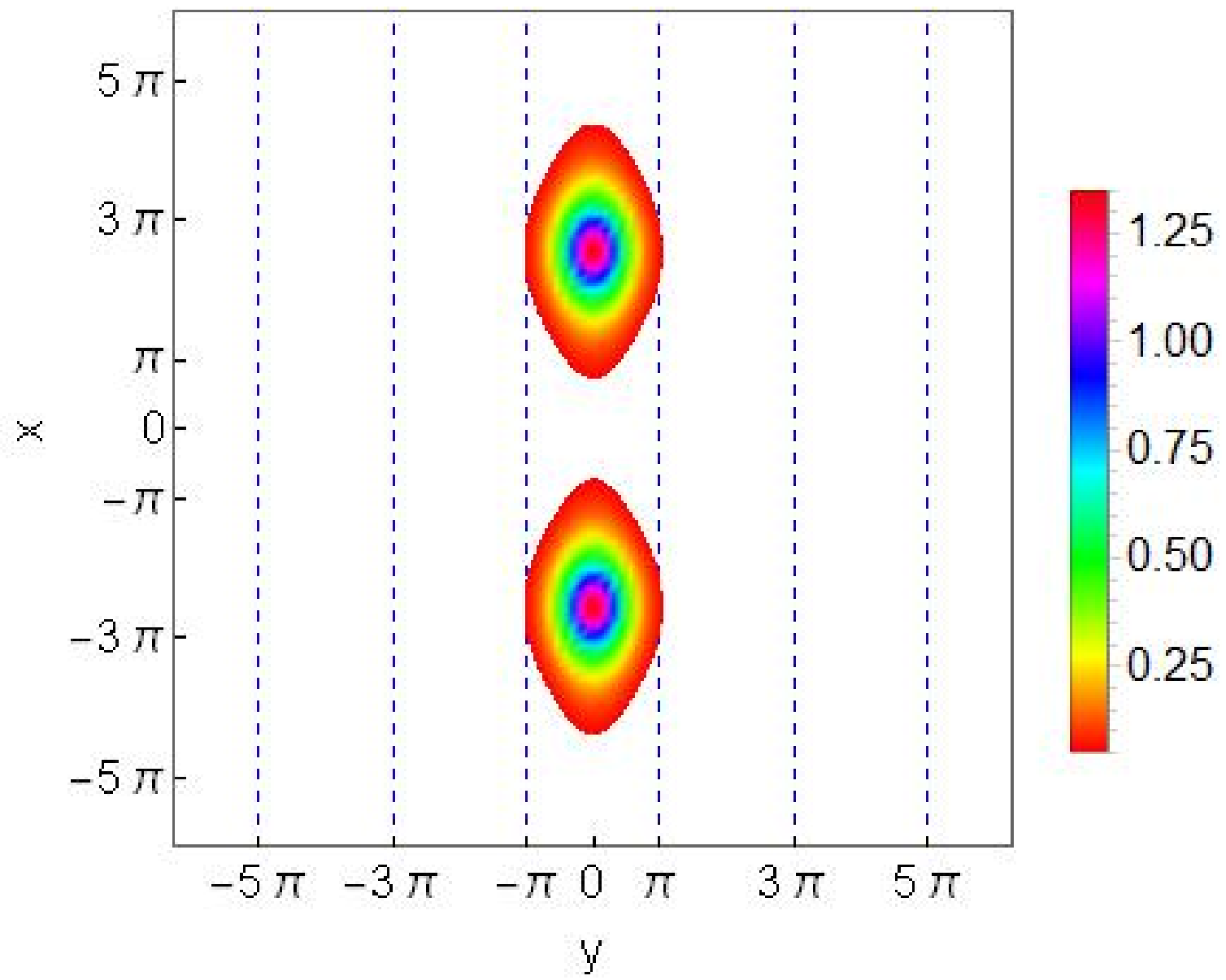}\quad
\includegraphics[width=4cm,height=4cm,clip]{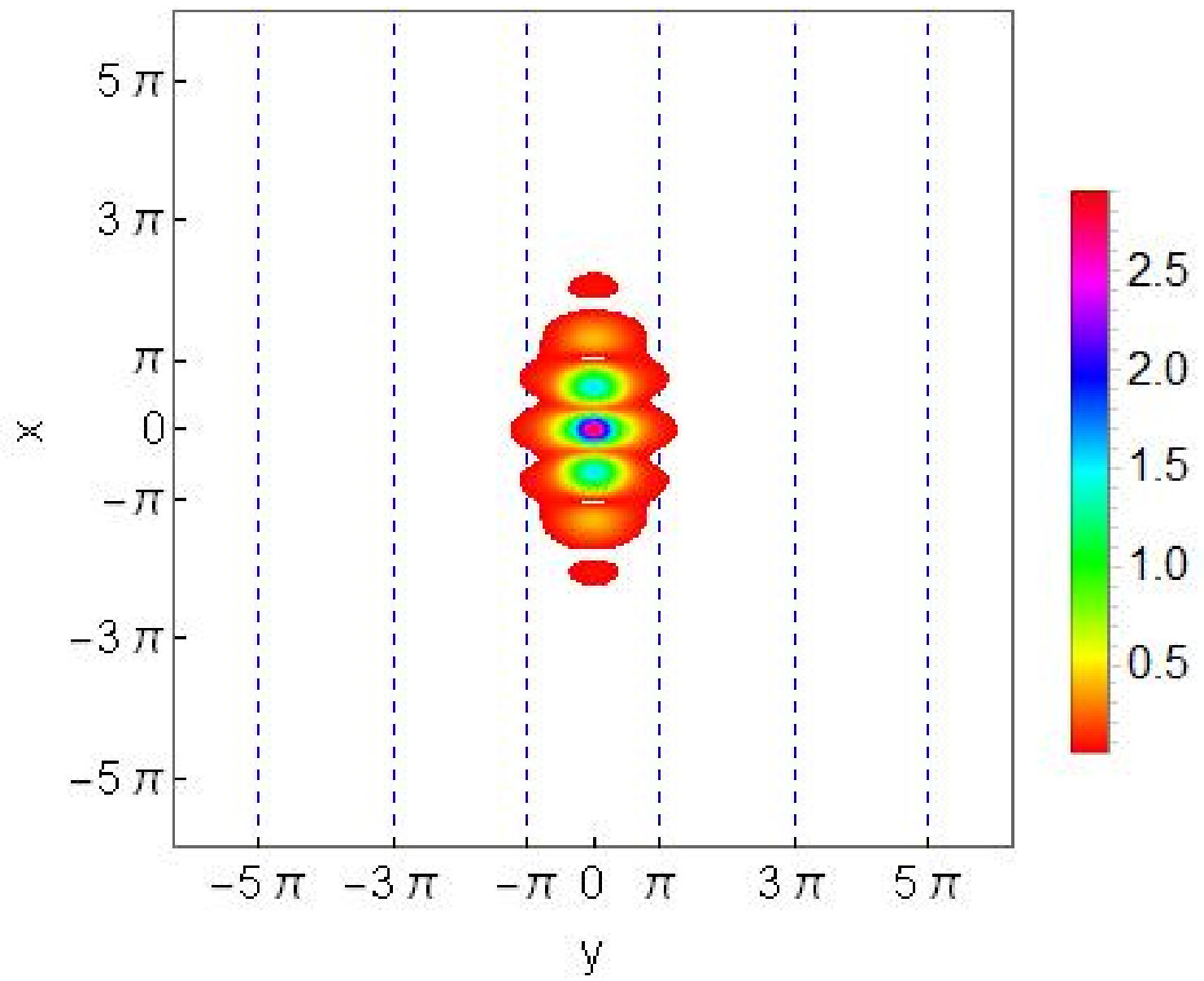}\quad
\includegraphics[width=4cm,height=4cm,clip]{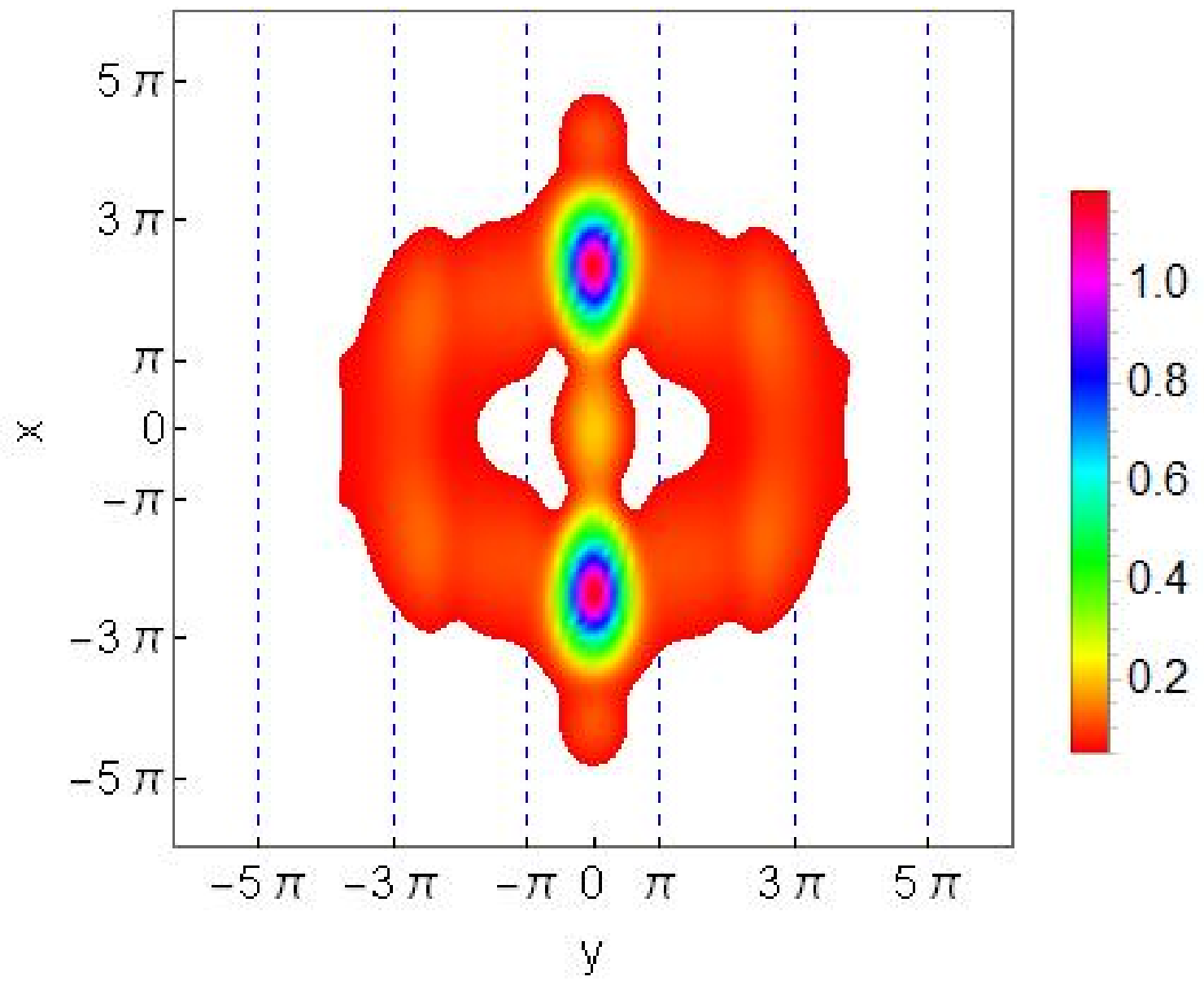}\quad
\includegraphics[width=4cm,height=4cm,clip]{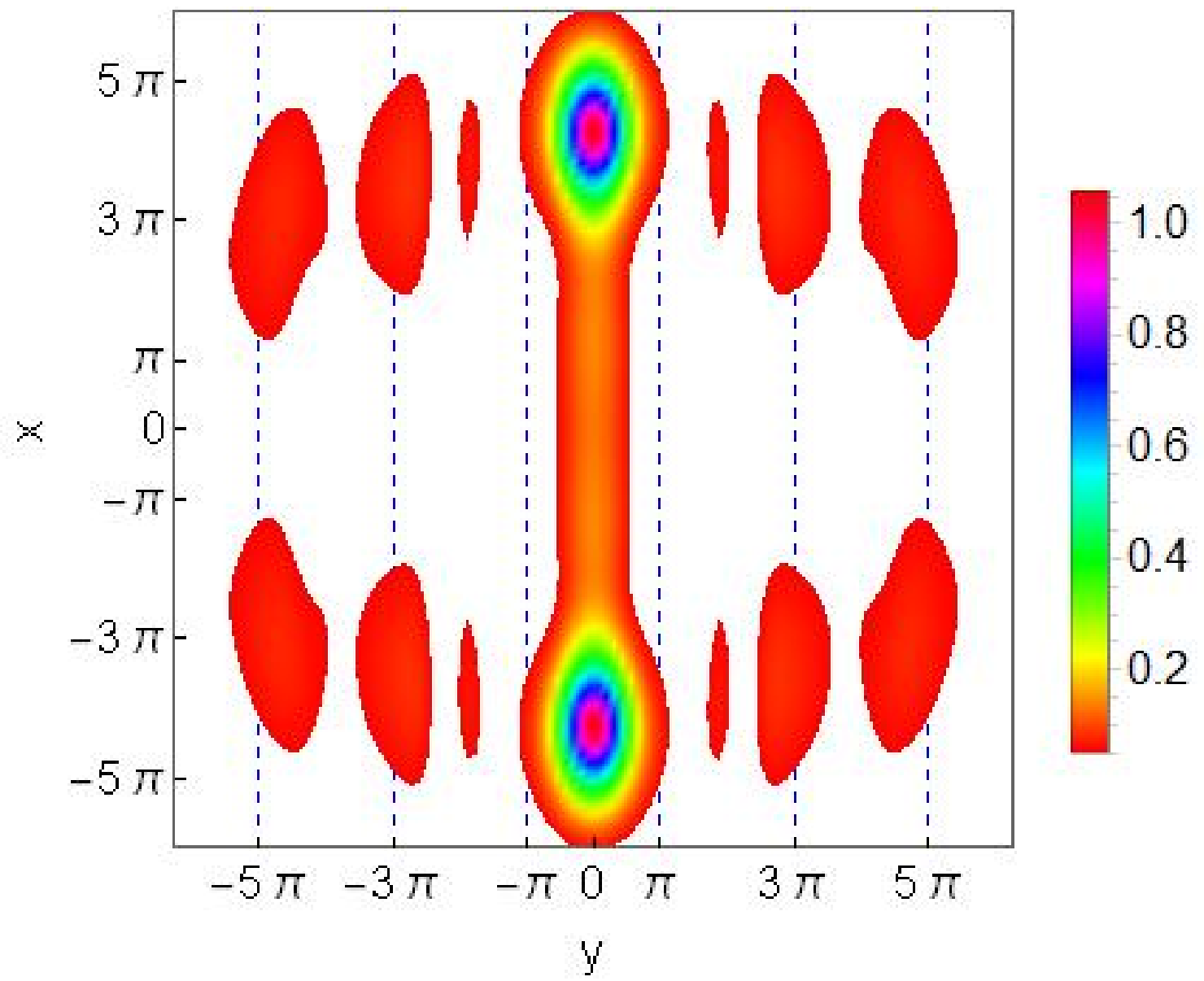}}
\caption{Collision of two SS moving in the same channel of the
quasi-1D OL. The solitons are set in motion towards each other by a
weak harmonic trap along the free direction  $ V(x,y) = 0.05 x^2 -
\cos(y)$. The vertical dashed lines indicate the maxima of the OL
potential. The time instances are shown above the figure panels.
Parameter settings are similar to previous figure. } \label{fig2}
\end{figure}

\begin{figure}[htb]
\centerline{t = 8 \hspace{3cm} t = 10 \hspace{3cm} t = 11
\hspace{3cm} t = 13 \quad} \vspace{0.2cm} \centerline{
\includegraphics[width=4cm,height=4cm,clip]{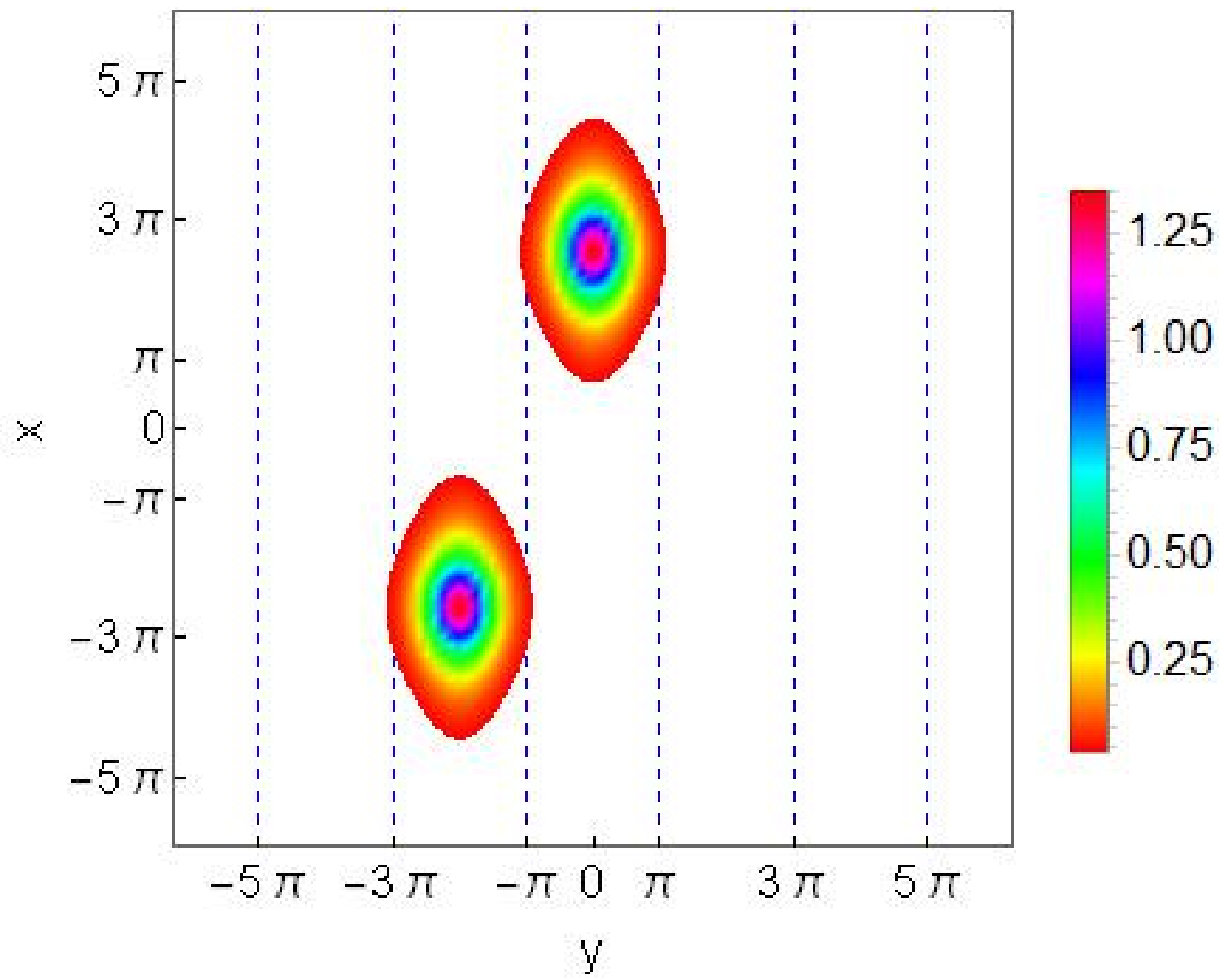}\quad
\includegraphics[width=4cm,height=4cm,clip]{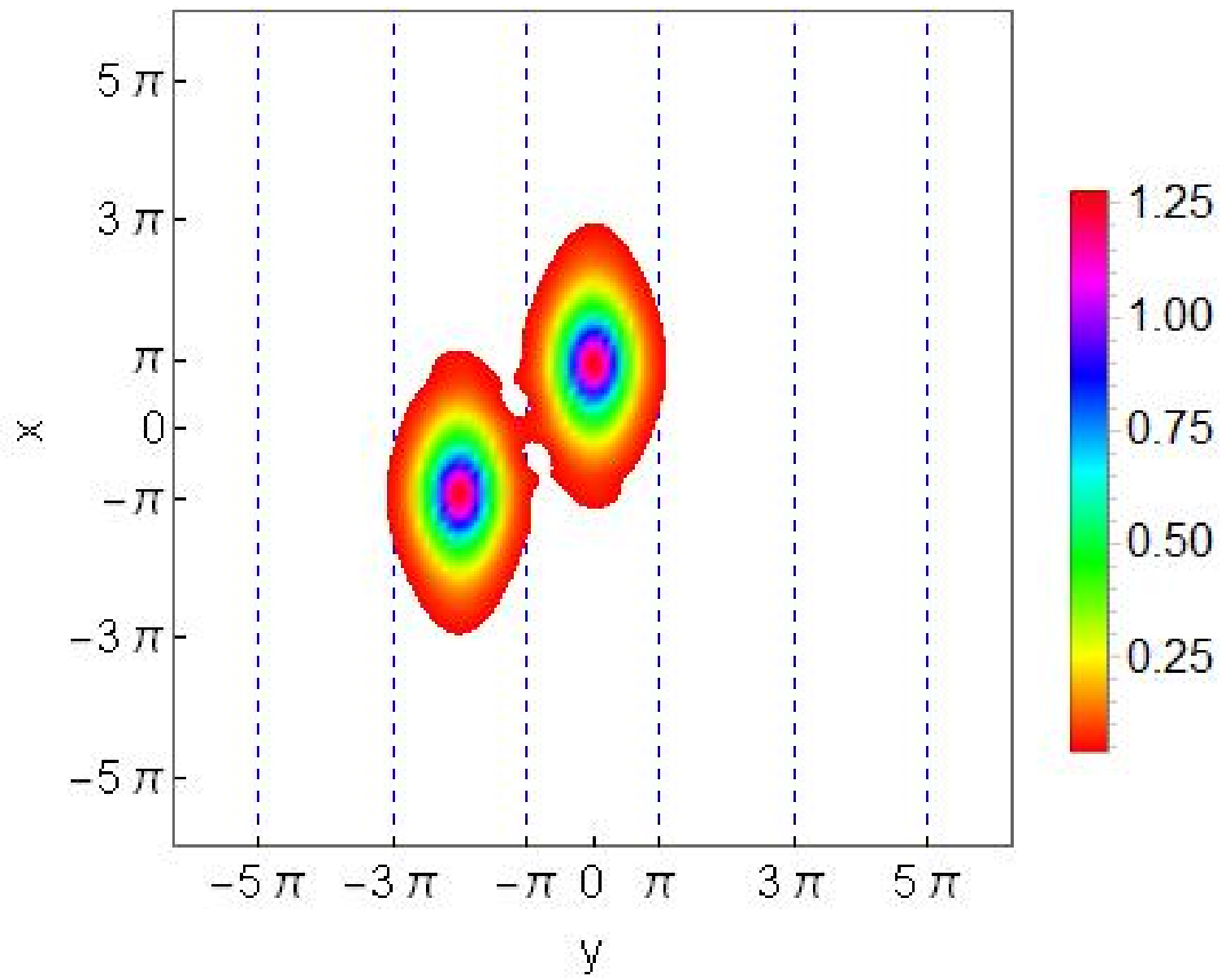}\quad
\includegraphics[width=4cm,height=4cm,clip]{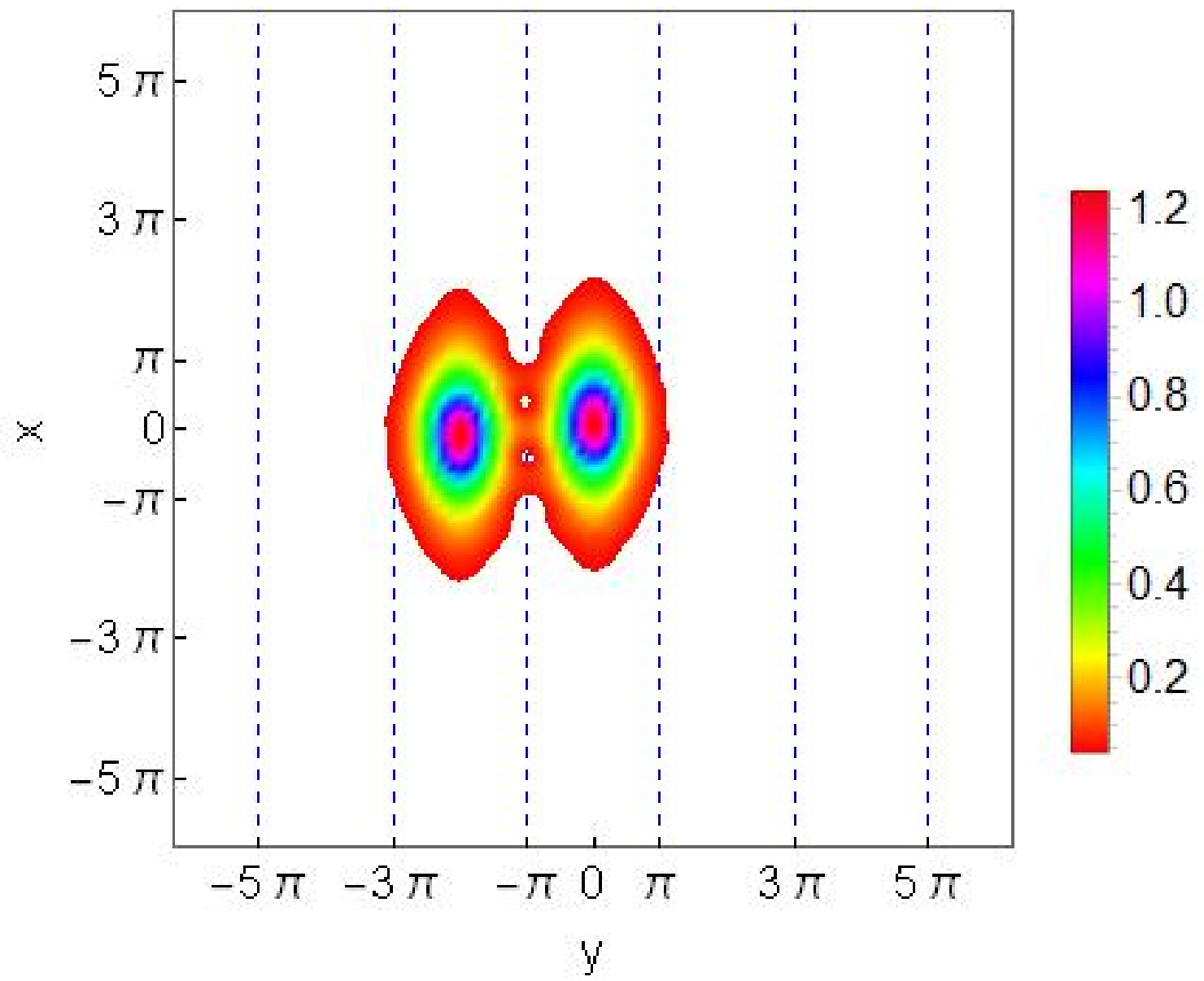}\quad
\includegraphics[width=4cm,height=4cm,clip]{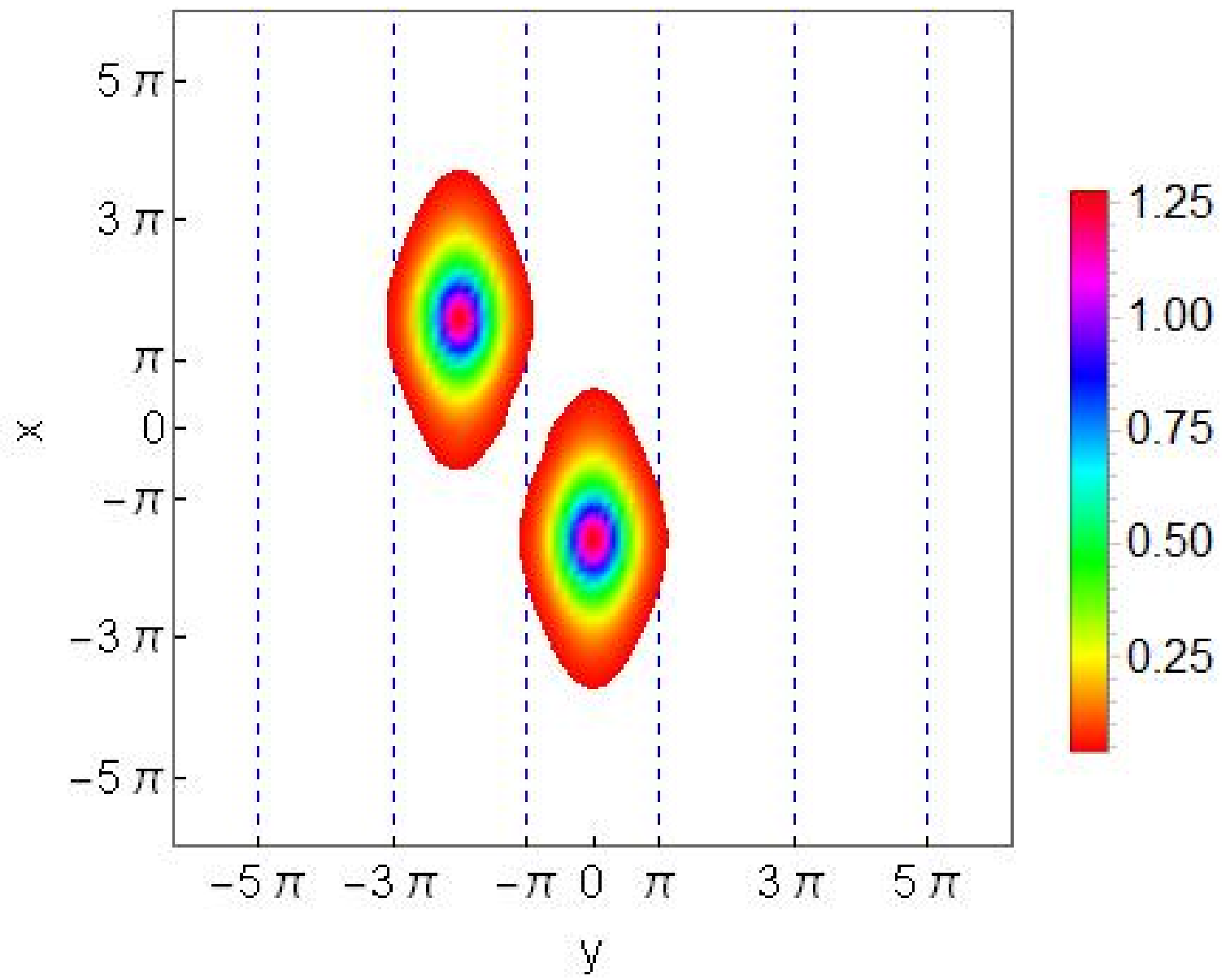}}
\caption{Collision of two SS moving in adjacent channels of the OL
potential. Solitons reappear unscathed after sideline collisions.
The time instances are shown above the figure panels. The parameter
values are similar to those in previous figure.} \label{fig3}
\end{figure}

Typical example of frontal collision of 2D symbiotic solitons moving
in a quasi-1D OL is illustrated in Fig. \ref{fig2}.
At the time of complete overlap ($t=11$), the central amplitude of
the localized state increases more than twice the initial value,
nevertheless, the wave collapse does not occur. Instead, the
collision produces small amplitude linear waves emitted by solitons,
which propagate in all directions. The periodic potential of the OL
appears to be transparent for propagating linear waves. When the two
solitons went through each other and then separated, they remain
connected by a small amplitude wave field ($t=17$). The density of
matter between solitons in the collision channel is small but almost
uniform, without interference fringes, which is evidence of its
incoherent nature. The wave field connecting the two outgoing SS
vanishes when they become well separated. Collision of two SS moving
in adjacent channels of the OL is shown in Fig. \ref{fig3}.
Under these circumstances solitons interact only through their small
amplitude tails and neither experience notable deformations nor emit
linear waves. We have not observed significant difference between
collisions of two in-phase and anti-phase SS. Interactions of 2D SS
appear to be less phase-sensitive compared to scalar 2D solitons.

\section{Quantum droplets in quasi-1D optical lattice}

Although quantum droplets are stable in free space, external
potentials can be helpful in revealing their peculiar liquid
properties, such as incompressibility and surface tension. For
instance, if the equilibrium shape of the droplet is first slightly
distorted by an external potential and then released, it will
oscillate with a frequency that is correlated to its surface
tension. Below we employ the quasi-1D OL potential to study
phenomena related to splitting and merging of quantum droplets. The
surface tension is expected to play a decisive role in these
processes.

It is well-known that the surface tension makes the droplet minimize
its interface with the surrounding medium (its surface area)
\cite{frohn-book}. In 3D free space, the equilibrium shape of the
droplet due to isotropy of space is a sphere, while in 2D space it
is a circle. The droplet subjected to squeezing in one direction
will then expand in the other (perpendicular) direction, and this is
a manifestation of its incompressibility. If the exerted force is
sufficiently strong, the droplet can split into fragments. The
number of fragments obeys the energy optimization. In particular,
splitting one 2D droplet into {\it two} equal fragments is
$\sqrt{3/2}$ times more energetically favorable than splitting it
into {\it three} equal fragments. Below we numerically investigate
splitting and merging phenomena with a 2D QD using the quasi-1D OL.

In contrast to symbiotic solitons considered in the previous
section, QD does not need any external potential for its stability.
Instead, the LHY quantum fluctuations term \cite{lee1957} included
in the governing GPE is sufficient for its complete stability.
However, we include both the LHY term and external potential in our
model, although the latter plays an auxiliary role. Thus the system
of coupled GPEs in scaled units has the form
\cite{chen2018,li2017,otajonov2022}
\begin{equation}\label{qd}
i\partial_t\psi_i+\Delta\psi_i-V(x,y)\psi_i+(g_i|\psi_i|^2+g_{ij}|\psi_j|^2)\,\psi_i
+ \gamma (|\psi_i|^2+|\psi_j|^2)\, {\rm ln}(|\psi_i|^2+|\psi_j|^2)\,
\psi_i = 0, \quad i,j = 1,2, \quad i \not= j,
\end{equation}
where $\gamma = f(g_{1},g_{2},g_{12})$ is the coupling constant
accounting for the quantum fluctuations in 2D space
\cite{petrov2016}, with imposed condition $g_{1} = g_{2}$. The
meanings of other terms are similar to those in Eqs. (\ref{gpe}).

\begin{figure}[htb]
\centerline{\quad a) \hspace{5cm} b) \hspace{4cm} c)} \centerline{
\includegraphics[width=5cm,height=4cm,clip]{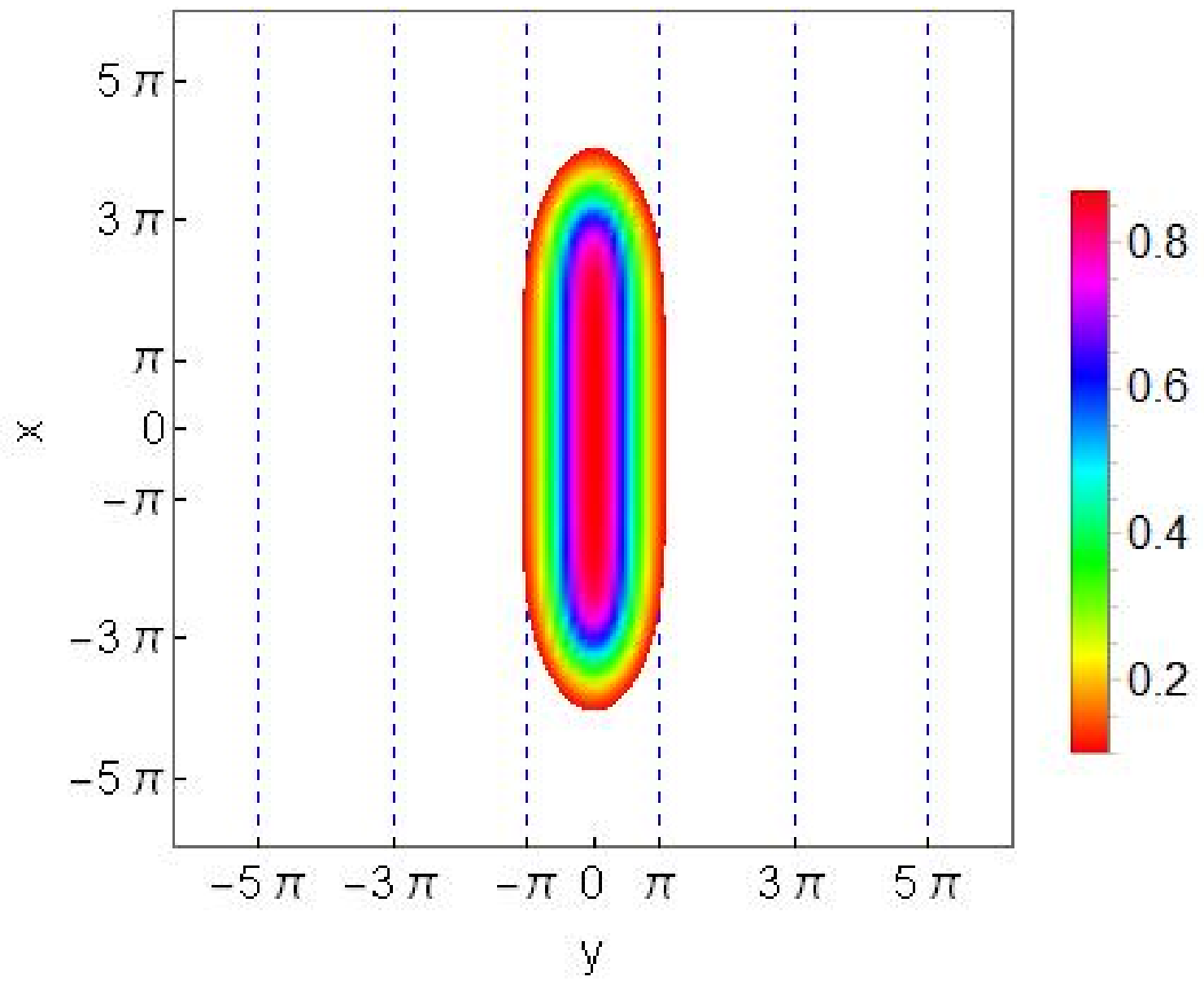}\quad
\includegraphics[width=5cm,height=4cm,clip]{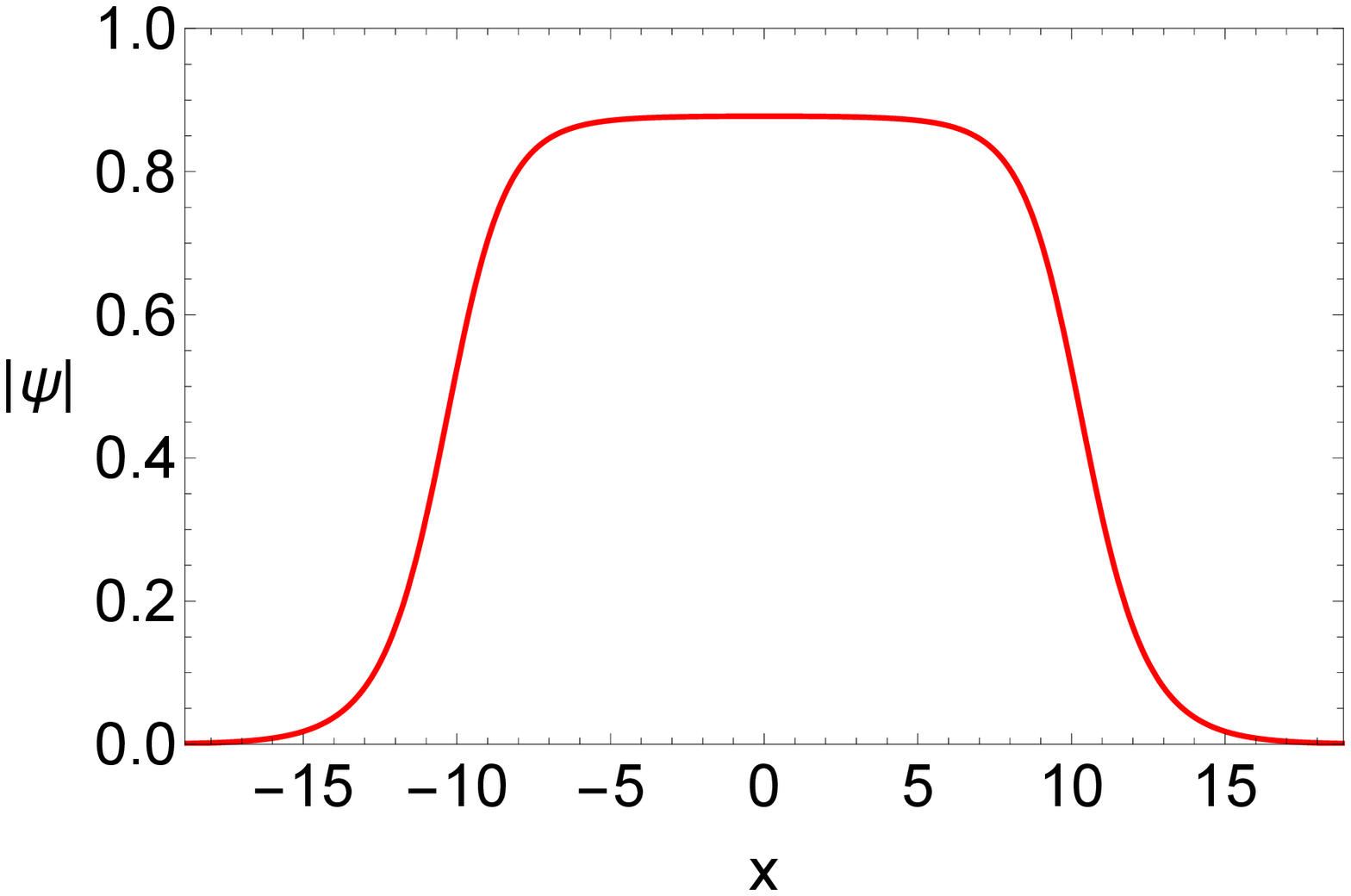}\quad
\includegraphics[width=5cm,height=4cm,clip]{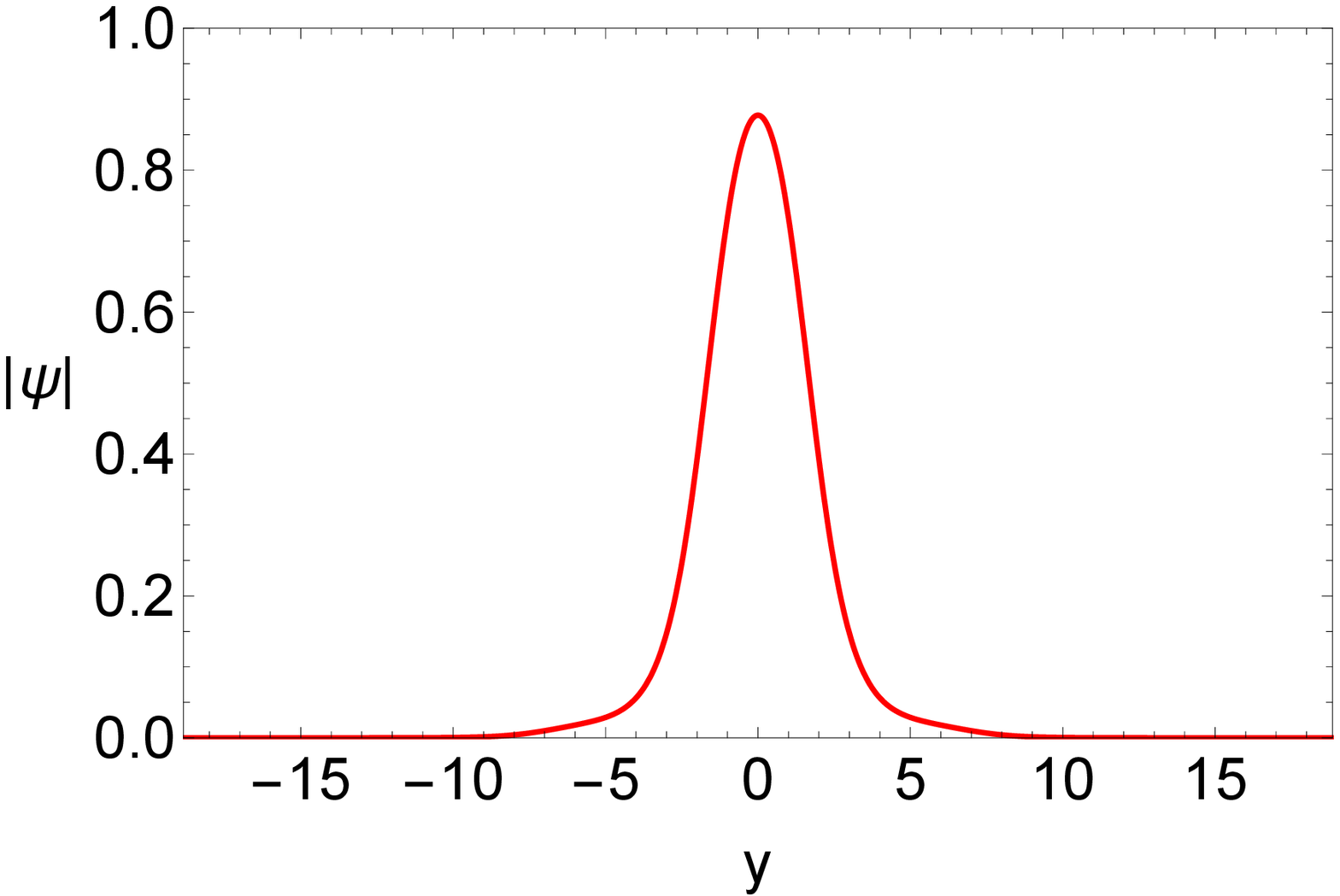}}
\caption{a) Density plot $|\psi(x,y)|$ for the 2D QD's ground state
in a quasi-1D optical lattice $V(x,y)=V_0 \cos(k y)$ according to
Eq. (\ref{qd2D}) obtained from initial waveform Eq. (\ref{ansatz})
with $A=1$, $a=4$, $b=10/\pi$. Vertical dashed lines indicate the
maxima of the OL potential. b) The wave profile $|\psi(x,0)|$ along
the free direction $x$ for $y=0$. c) The wave profile $|\psi(0,y)|$
along the $y$-axis for $x=0$. Parameter values: $V_0 = -0.6$, $k=1$,
$N = 40$. } \label{fig4}
\end{figure}

In this section we consider parameter settings for a symmetric two
component 2D BEC $\psi_1 = \psi_2 =\psi$, with equal atom numbers of
the components $N_1 = N_2 = N$, equal intra-species coupling
coefficients $g_{1}=g_{2}$ and $g_{12} \simeq -\sqrt{g_{1} g_{2}}$.
Under these conditions and additional scaling $\psi \rightarrow
\psi/\sqrt{2}$ the system of coupled GPE's (\ref{qd}) reduces to a
single component GPE
\begin{equation}\label{qd2D}
i\psi_t+\Delta \psi - V(x,y)\psi - 2 \,|\psi|^2 {\rm ln}(|\psi|^2)
\psi = 0.
\end{equation}
In the first stage of our numerical experiment we produce a 2D QD in
a quasi-1D OL using the Pitaevskii phenomenological damping
procedure \cite{choi1998}, as illustrated in Fig. \ref{fig4}. The
apparent flat-top shape of the 2D QD in the unconstrained direction
$x$ of the potential indicates that it is in an incompressible
regime \cite{ferioli2019}. The surface tension tends to expand the
droplet in the $y$ direction against the squeezing action of the OL
potential. If the potential is sufficiently weak the surface tension
can prevail, leading to its expansion and/or splitting.

\begin{figure}[htb]
\centerline{\quad a) \hspace{5cm} b) \hspace{4cm} c)} \centerline{
\includegraphics[width=5cm,height=4cm,clip]{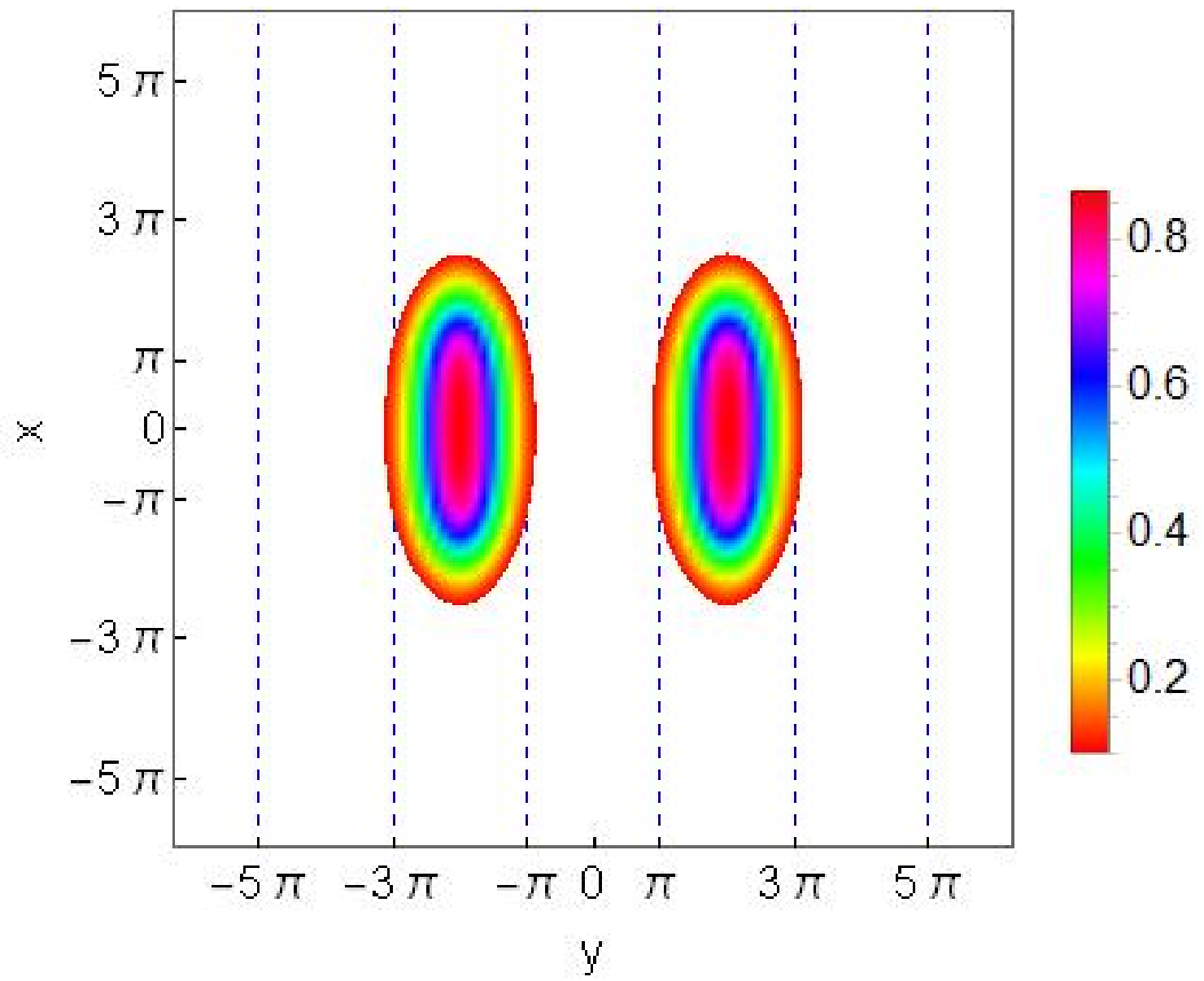}\quad
\includegraphics[width=5cm,height=4cm,clip]{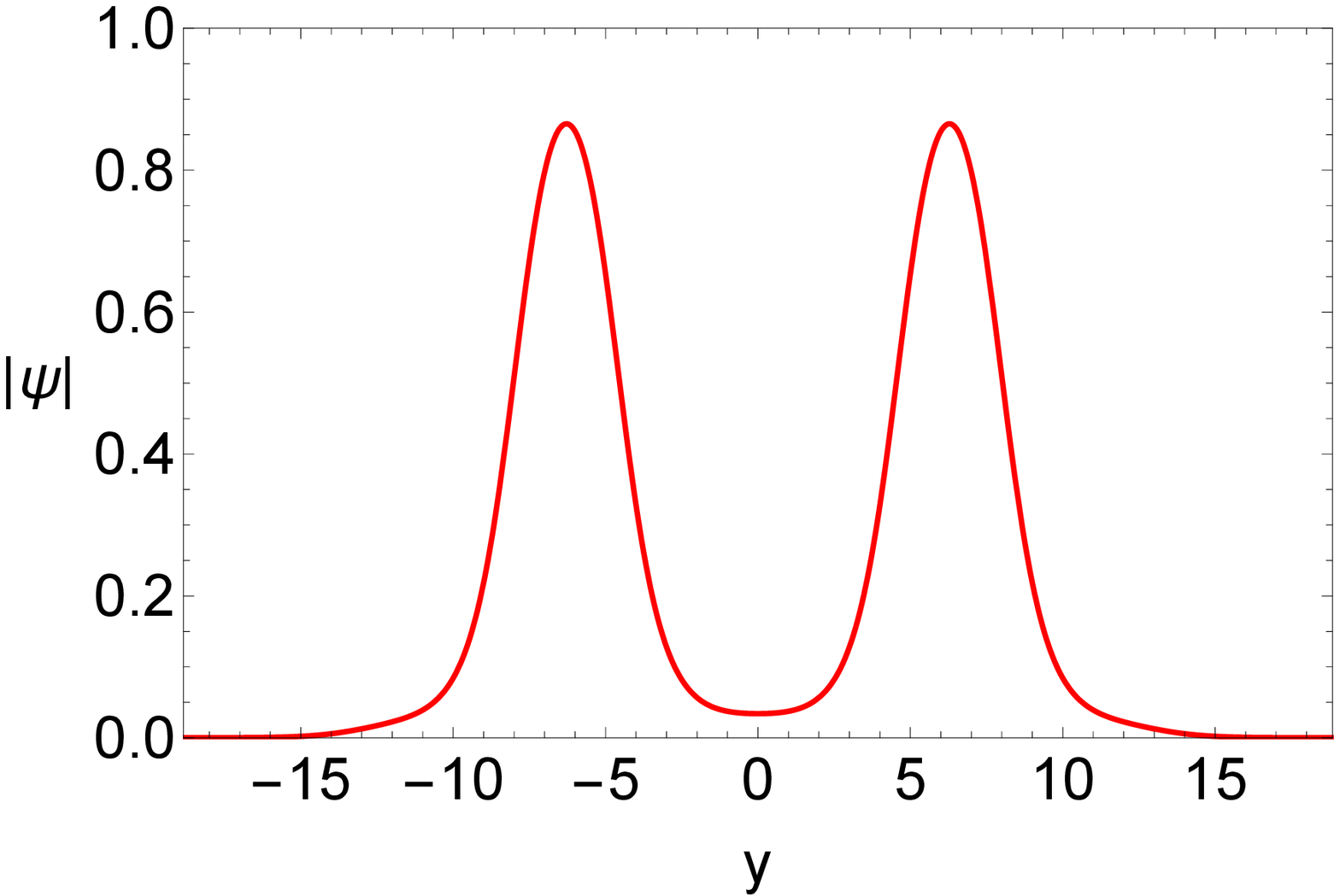}\quad
\includegraphics[width=5cm,height=4cm,clip]{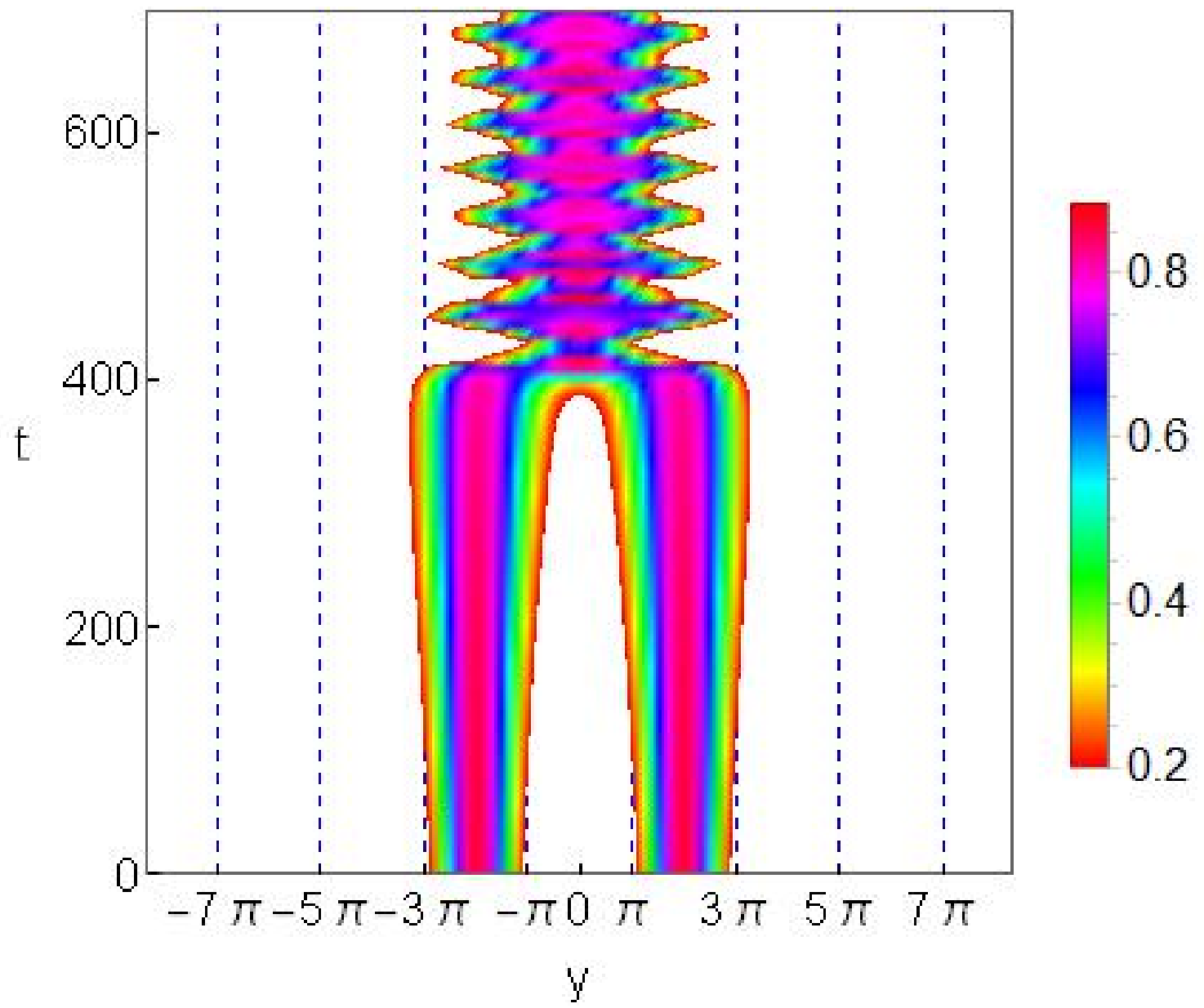}}
\caption{a) The ground state for weaker OL potential is the
fragmented QD, shown through the density plot. b) The cross section
profile $|\psi(0,y)|$ of the fragmented QD along the $y$-axis. A
weak link remains between the fragments. c) Time evolution of the
wave profile $|\psi(0,y,t)|$ when the confining potential is slowly
reduced to zero according to law $V(y,t) = (1-{\rm tanh}(3\,
t/t_{end}))\cdot V_0 \cos(y)$ with, $t_{end} = 1000$. At $t \sim
400$ the fragments merge in a manner, which is typical for the
surface tension driven coalescence of two liquid drops, without
emission of linear matter waves. Parameter values are similar to
those in the previous figure, except $V_0=-0.5$. A factor 3 in ${\rm
tanh}(.)$ is used for the smooth transition.} \label{fig5}
\end{figure}

In the second stage, we use an OL of reduced strength to allow the
droplet to expand/split in the $y$ direction. The ground state
produced by the Pitaevskii damping procedure \cite{choi1998} for the
weaker OL is shown in Fig. \ref{fig5} a,b. What is peculiar about
this splitting process is that the QD splits into two similar
fragments, leaving its original location (cell of OL at $y=0$)
empty, in accordance with the above-mentioned energy considerations.

In the final third stage of our numerical experiment, we explore the
merging process of QD fragments. We believe that surface tension
should drive the process of coalescence of two similar droplets into
a single drop with a smaller surface area (length of the circle in
2D). To initiate the dynamics of this merging process we slowly
switch off the OL and leave the droplets under their own mutual
interactions. In free space, the droplets acquire a circular shape
(initially distorted by a quasi-1D OL) by expanding in the $y$
direction, while the weak link remaining between the fragments
(Fig.~\ref{fig5} b) strengthens. The weak link (bridge of nonzero
matter density) is provided by a suitable period and strength of the
quasi-1D OL, as well as the norm (number of atoms) of the QD.

The process of merging two QD fragments shown in Fig. \ref{fig5}\,c
bears the signatures of coalescence of liquid droplets driven by
surface tension. At first, the weak link (thin neck) between the
droplets becomes stronger while the droplets expand toward each
other. When the coalescing droplets come into contact the process of
their merging occurs extremely rapidly similarly to what happened
with classical liquid droplets \cite{eggers1999,duchemin2003}. The
resulting single droplet oscillates without the emission of linear
matter waves, which can be regarded as another manifestation of the
liquid nature of the QD. In contrast, usual solitonic wave packets
emit radiation when perturbed \cite{novikov-book}.

\section{Quantum droplets in imbalanced Bose-gas mixtures}

Imbalanced Bose-gas mixtures where the number of atoms in one
component exceeds the number of atoms in the other component
represent a convenient system for exploring the properties of QDs
immersed in a superfluid (SF) environment (residual condensate). One
of the relevant problems concerns the investigation of anisotropic
speed of sound in a density-modulated superfluid BEC, recently
reported in \cite{tao2023,chauveau2023}. In these works the authors
created periodic modulation of the density in BEC using 1D OL. The
speed of sound, and therefore SF density was found to be reduced
along the direction of OL. In the model considered below the motion
of a QD through the density-modulated residual condensate along the
free direction and across the OL may create sound waves of different
velocities, thus revealing associated SF anisotropy.

In imbalanced mixtures the excess atoms of the larger component that
cannot bind to the droplet must be confined by external potential or
by imposing periodic boundary conditions. In experiments, the
periodic boundary conditions are realized in toroidal traps
\cite{ramanathan2011,sherlock2011}. The existence of QDs in binary
Bose-gas mixtures with equal intra-component interactions but
unequal number of atoms in the two components in a one-dimensional
setting was recently reported in~\cite{tengstrand2022}. It should be
emphasized that the formation of QDs in binary condensates with
inter-component asymmetry is a problem of fundamental interest. In
this regard, by examining the QD's rotation through the background
condensate in a ring trap and emerging collective excitations it was
revealed that this compound system can exhibit both rigid-body and
superfluid properties \cite{tengstrand2022}.

In this section, we consider imbalanced binary 2D BEC in the
presence of a quasi-1D OL. Due to the inter-species attraction the
bigger component can act as a trapping medium for the smaller one.
Therefore trapping only the bigger component by OL and imposing
periodic boundary conditions is sufficient for exploring the
dynamics of QDs in imbalanced Bose-gas mixtures.

\begin{figure}[htb]
\centerline{ a) \hspace{4cm} b) \hspace{4cm} c) \hspace{4cm} d)}
\centerline{
\includegraphics[width=4cm,height=4cm,clip]{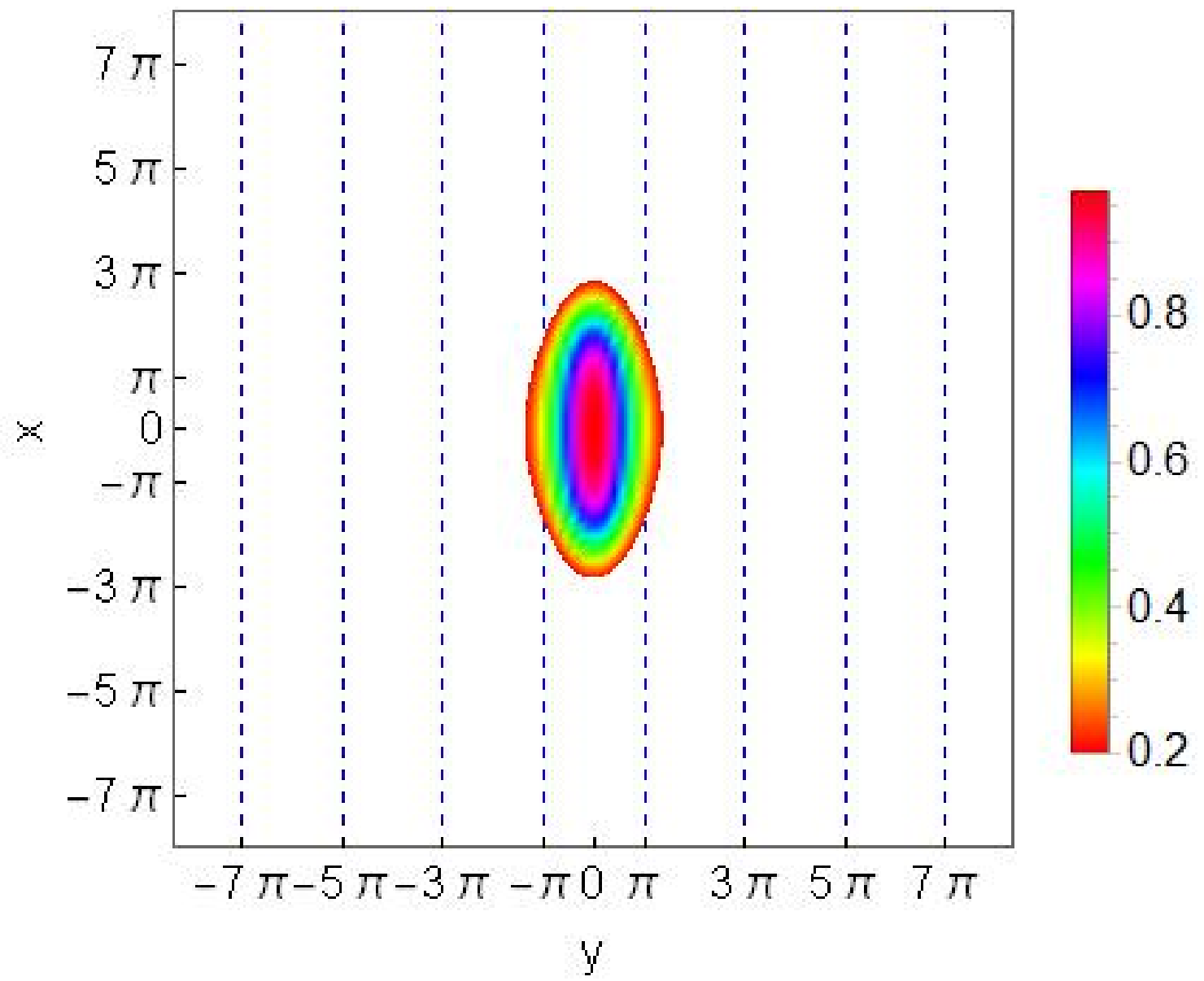}\quad
\includegraphics[width=4cm,height=4cm,clip]{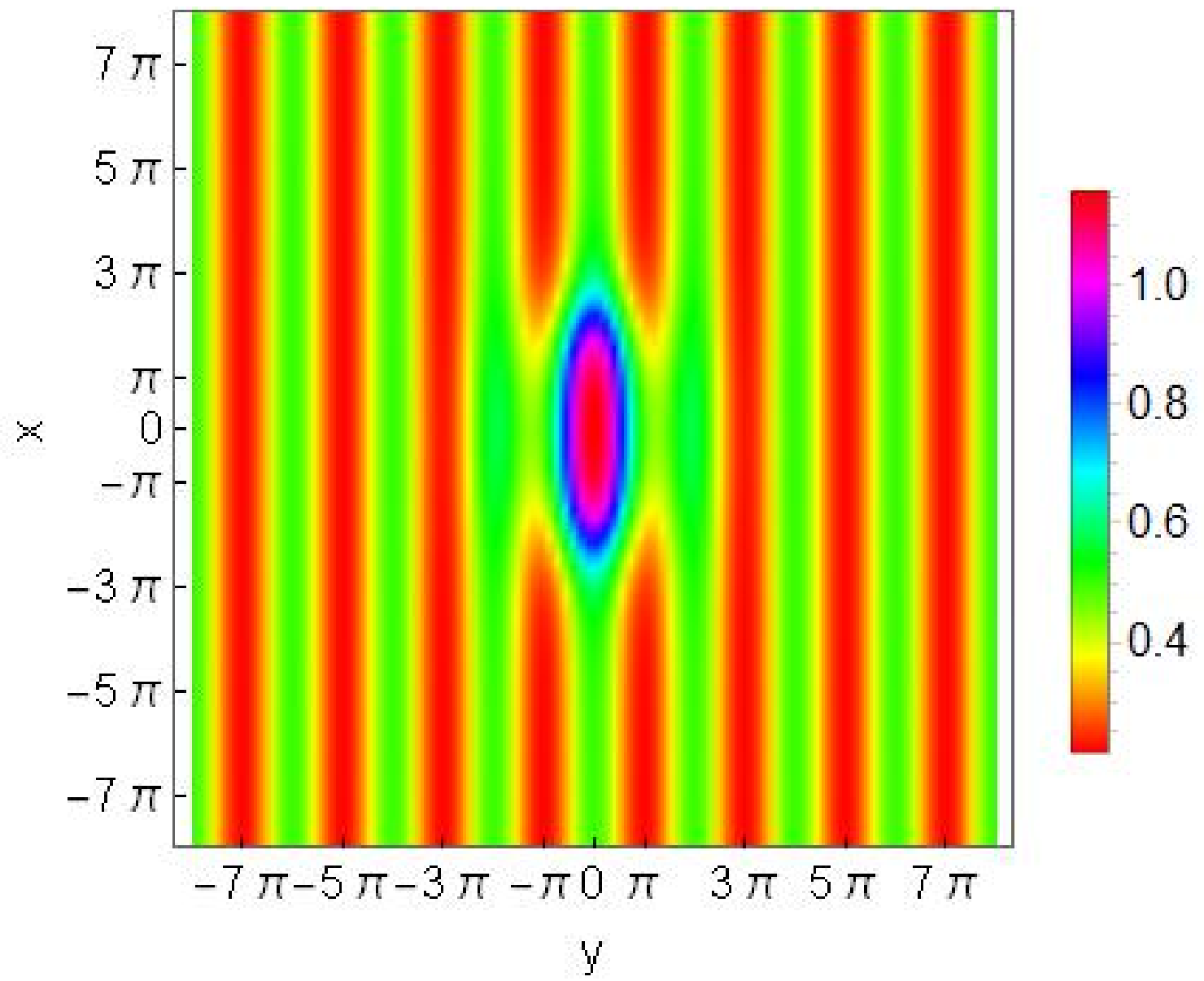}\quad
\includegraphics[width=4cm,height=4cm,clip]{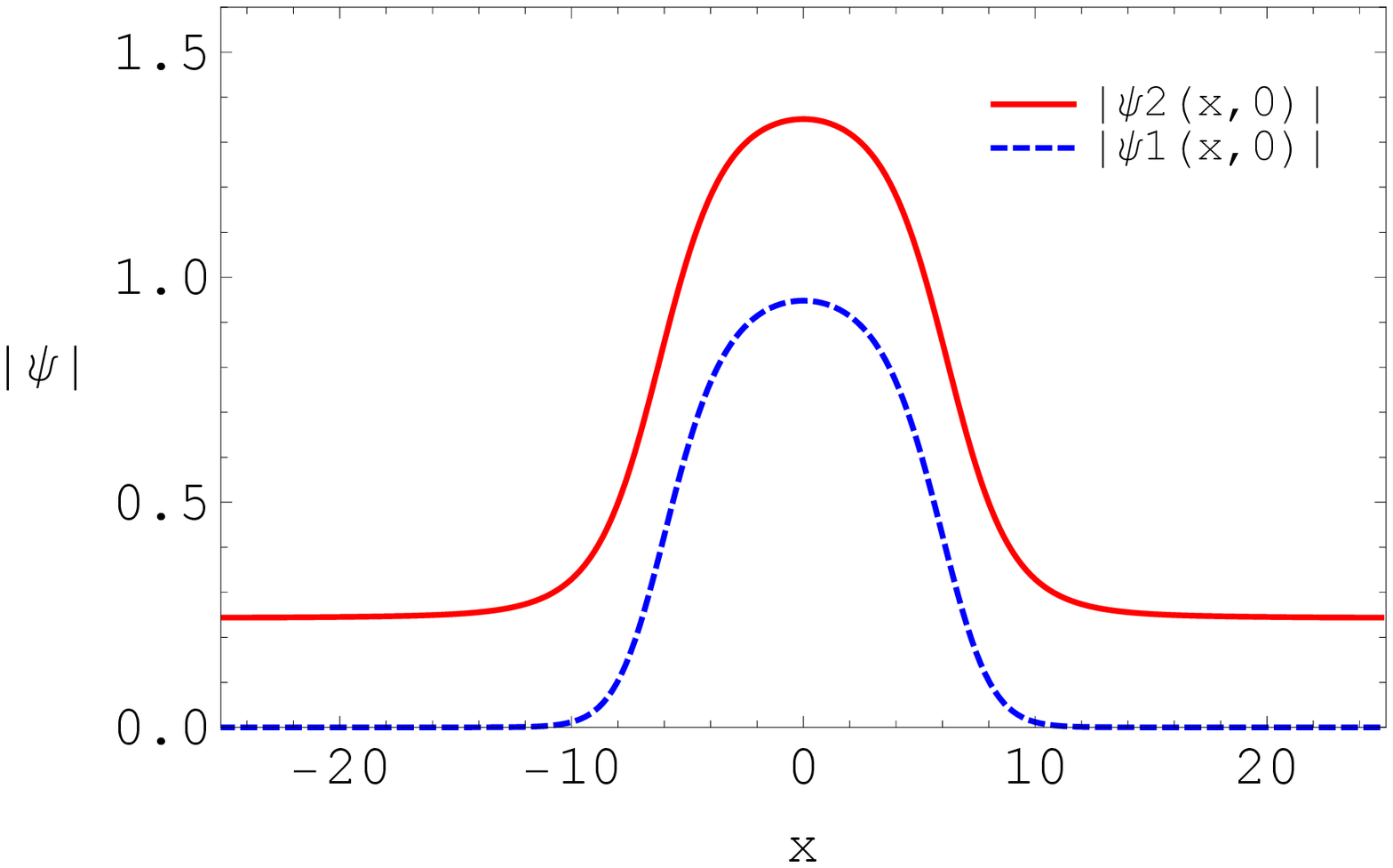}\quad
\includegraphics[width=4cm,height=4cm,clip]{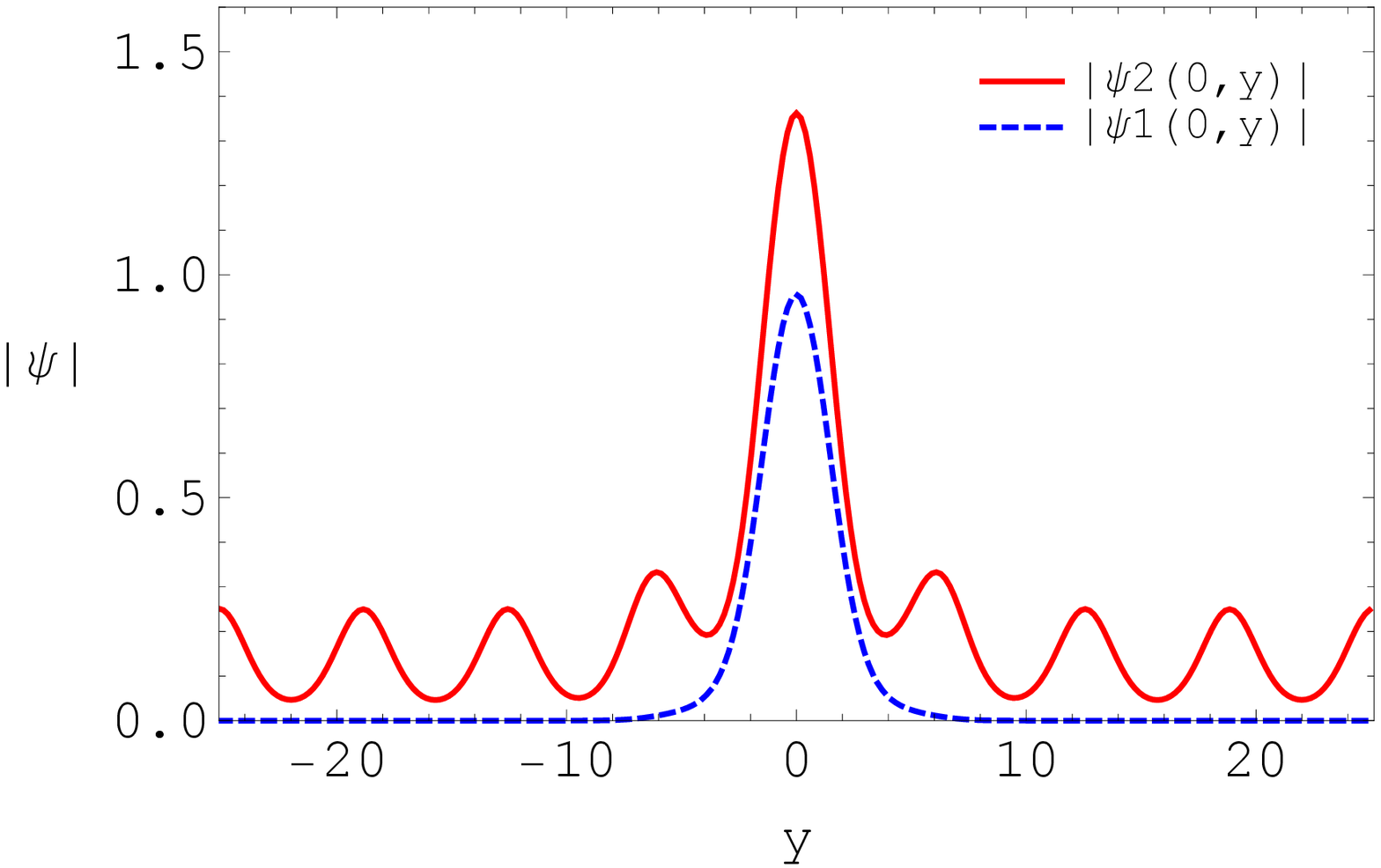}}
\caption{(a,b) Density plots for the smaller ($|\psi_1|$) and bigger
($|\psi_2|$) components of the binary BEC according to Eq.
(\ref{qd}). (c,d) The cross-section wave profiles along the free
direction at $y=0$ and across the quasi-1D OL at $x=0$. Parameter
values: $N_1 = 40$, $N_2 = 400$, $g_1 = g_2 = -1$, $g_{12} = 2$,
$\gamma = -0.4$. The potential $V(x,y) = V_0 \cos(y)$ with $V_0 =
-0.5$ acts only upon the larger component~$\psi_2$.} \label{fig6}
\end{figure}

The ground state wave profiles are obtained using the Pitaevskii
damping procedure \cite{choi1998}. As initial conditions, we employ
the trial functions given by Eq.(\ref{ansatz}) with appropriate
parameters for the amplitude and widths that correspond to the given
norms of the components. A typical result is shown in Fig.
\ref{fig6} where the smaller component occupy a single cell of the
quasi-1D OL, while the larger component is distributed over the
whole domain with periodically modulated density along $y$
direction.

The QD immersed in the background condensate of the larger component
can move in both directions. When it moves along the free direction
$x$ of the potential, it experiences the background condensate with
uniform density, while moving in the orthogonal direction it
experiences the medium with periodically modulated density. Since
the motion of a probe through the superfluid can generate sound
waves, the considered model can be of interest to studies related to
anisotropic sound velocity in superfluids.

\section{Conclusions}

The properties of 2D symbiotic solitons and quantum droplets loaded
on a quasi-1D optical lattice have been explored. We have
demonstrated that the quasi-1D OL provides complete stability of 2D
SS against collapse or decay while it helps to illustrate
manifestations of the incompressibility and surface tension of QDs.
The mobility of SSs in the free direction of the quasi-1D OL allowed
us to investigate their frontal and sideline collisions. In
numerical experiments, we observed the splitting and merging of QDs
which resembles the behavior of classical liquid droplets. In
particular, when the two droplets come into contact their
coalescence occurs on an extremely fast time scale and the resulting
single droplet oscillates without emission of linear matter-waves
which reflects the preservation of its norm. The ground state wave
profiles are obtained for the imbalanced binary Bose-gas mixtures
with intra-species repulsion and inter-species attraction in
presence of a quasi-1D OL. The smaller component, which is a fully
bound QD, can freely move through the residual large component of
the condensate and excite sound waves in it. The proposed model can
be helpful in exploring the superfluidity in density-modulated
media.

\section{Acknowledgments}

All authors gratefully acknowledge the support provided by
Interdisciplinary Research Center for Intelligent Secure Systems
(IRC-ISS) at King Fahd University of Petroleum and Minerals (KFUPM)
in funding this work through project No. INSS2302. BBB thanks the
Physics Department at KFUPM for their hospitality during his visit.

\end{document}